\documentclass[fleqn,11pt,twoside]{article}

\usepackage{amsthm,amsthm,amssymb, color, xcolor,epsfig, graphics, subfigure}

\usepackage{amsmath, graphicx, latexsym, lscape }

\makeatletter
\newcommand{\copyrightnote}[2]{{\renewcommand{\thefootnote}{}
 \footnotetext{\small\it
\begin{flushleft}
 \copyright \ #1   #2  
\end{flushleft}}}}

\newcommand{\Name}[1]{\begin{flushleft}
                       \LARGE \bf #1
                       \end{flushleft}\vspace{-3mm}}

\newcommand{\Author}[1]{\begin{flushleft}
                       \it #1 \end{flushleft}}

\newcommand{\Address}[1]{\begin{flushleft}
                       \it #1 \end{flushleft}}

\newcommand{\Date}[1]{\begin{flushleft}
                      \small  \it #1 \end{flushleft}}

\newcommand{\evenhead}{Author \ name}
\newcommand{\oddhead}{Article \ name}

\renewcommand{\@evenhead}{
\hspace*{-3pt}\raisebox{-15pt}[\headheight][0pt]{\vbox{\hbox to \textwidth
{\thepage \hfil \evenhead}\vskip4pt \hrule}}}
\renewcommand{\@oddhead}{
\hspace*{-3pt}\raisebox{-15pt}[\headheight][0pt]{\vbox{\hbox to \textwidth
{\oddhead \hfil \thepage}\vskip4pt\hrule}}}
\renewcommand{\@evenfoot}{}
\renewcommand{\@oddfoot}{}

\long\def\@makecaption#1#2{%
  \vskip\abovecaptionskip
  \sbox\@tempboxa{\small \textbf{#1.}\ \ #2}%
  \ifdim \wd\@tempboxa >\hsize
    {\small \textbf{#1.}\ \ #2}\par
  \else
    \global \@minipagefalse
    \hb@xt@\hsize{\hfil\box\@tempboxa\hfil}%
  \fi
  \vskip\belowcaptionskip}

\newcommand{\JNMPnumberwithin}[3][\arabic]{%
  \@ifundefined{c@#2}{\@nocounterr{#2}}{%
    \@ifundefined{c@#3}{\@nocnterr{#3}}{%
      \@addtoreset{#2}{#3}%
      \@xp\xdef\csname the#2\endcsname{%
        \@xp\@nx\csname the#3\endcsname .\@nx#1{#2}}}}%
}

\renewenvironment{proof}[1][\proofname]{\par
  \normalfont
  \topsep6\p@\@plus6\p@ \trivlist
  \item[\hskip\labelsep\textbf{%
    #1\@addpunct{.}}]\ignorespaces
}{%
  \qed\endtrivlist
}

\newcommand{\resetfootnoterule} {
  \renewcommand\footnoterule{%
  \kern-3\p@
  \hrule\@width.4\columnwidth
  \kern2.6\p@}
}

\renewcommand{\footnoterule}{}

\makeatother

\theoremstyle{definition}

\theoremstyle{plain}
\newtheorem{theorem}{Theorem}[section]

\newtheorem{proposition}[theorem]{Proposition}

\theoremstyle{definition}
\newtheorem{remark}[theorem]{Remark}

\newcommand{\sub}[1]{_{\mathrm{#1}}}
\newcommand{\su}[1]{^{\mathrm{#1}}}

\newcommand{\Id}{\mathbf{1}}
\newcommand{\eu}{\mathrm{e}}
\newcommand{\iu}{\mathrm{i}}
\newcommand{\di}{\mathrm{d}}
\newcommand{\N}{\mathbb{N}}
\newcommand{\Z}{\mathbb{Z}}

\newcommand{\R}{\mathbb{R}}
\newcommand{\C}{\mathbb{C}}

\newcommand{\cB}{\mathcal{B}}
\newcommand{\cD}{\mathcal{D}}
\newcommand{\cM}{\mathcal{M}}
\newcommand{\cH}{\mathcal{H}}
\newcommand{\cE}{\mathcal{E}}
\newcommand{\fa}{\mathfrak{a}}
\newcommand{\norm}[1]{\left\| #1 \right\|}
\newcommand{\scal}[2]{\left\langle #1, #2 \right\rangle}

\newcommand{\set}[1]{ \left\{  #1 \right\}} 

\DeclareMathOperator{\re}{Re}
\DeclareMathOperator{\im}{Im}
\DeclareMathOperator{\Span}{Span}

\begin{document}

\renewcommand{\evenhead}{ {\LARGE\textcolor{blue!10!black!40!green}{{\sf \ \ \ ]ocnmp[}}}\strut\hfill  M Ferrero and D Monaco}
\renewcommand{\oddhead}{ {\LARGE\textcolor{blue!10!black!40!green}{{\sf ]ocnmp[}}}\ \ \ \ \  
Effective quantum dynamics for magnetic fermions}

\thispagestyle{empty}
\newcommand{\FistPageHead}[3]{
\begin{flushleft}
\raisebox{8mm}[0pt][0pt]
{\footnotesize \sf
\parbox{150mm}{{Open Communications in Nonlinear Mathematical Physics}\ \  {\LARGE\textcolor{blue!10!black!40!green}{]ocnmp[}}
\ Vol.4 (2024) pp
#2\hfill {\sc #3}}}\vspace{-13mm}
\end{flushleft}}

\FistPageHead{1}{\pageref{firstpage}--\pageref{lastpage}}{ \ \ Article}

\strut\hfill

\strut\hfill

\copyrightnote{The author(s). Distributed under a Creative Commons Attribution 4.0 International License}

\Name{Effective quantum dynamics for magnetic fermions}

\Author{Margherita Ferrero}

\Address{Mathematisches Institut, Ludwig-Maximilians Universit\"{a}t M\"{u}nchen, Theresienstra{\ss}e 39, 80333 M\"{u}nchen, Germany}

\Author{Domenico Monaco}

\Address{Dipartimento di Matematica, Sapienza Universit\`{a} di Roma, Piazzale Aldo Moro 5, 00185 Roma, Italy}

\Date{Received June 24, 2024; Accepted September 5, 2024}

\setcounter{equation}{0}

\begin{abstract}
\noindent 
We show how to derive an effective nonlinear dynamics, described by the Hartree--Fock equations, for fermionic quantum particles confined to a two-dimensional box and in presence of an external, uniform magnetic field. The derivation invokes the Dirac--Frenkel principle. We discuss the validity of this effective description with respect to the many-body Schr\"odinger dynamics for small times and for weak interactions, and also in regards to the number of particles.
\end{abstract}

\label{firstpage}

\section{Introduction} \label{sec:Intro}

Many-body quantum theory is plagued by the ``curse of dimensionality'': although the many-body Schr\"odinger dynamics is described mathematically by a linear PDE, whose solution is therefore immediate to derive, the presence of a large number of interacting particles makes the concrete computation of properties of the solution (say, measured in terms of expectation values of observables in the many-body wave function) practically inaccessible. One looks therefore for \emph{effective descriptions} of the many-body quantum dynamics, which retain only partial (but still relevant) information on the full wave function, but on the other hand have the advantage of depending on fewer degrees of freedom. The ``price to pay'' to obtain this simplified representation is that the approximating objects are often described as solutions to \emph{nonlinear} differential equations. There are many instances of nonlinear PDEs of relevance for mathematical physics which arise in this way: from the nonlinear Schr\"odinger equation, used to model Bose--Einstein condensates~\cite{PitaevskiiStringari03}, to the Hartree (respectively Hartree--Fock) equations for the effective dynamics of large collections of bosons (respectively fermions)~\cite{Lubich08}. The derivation of such equations from microscopic models, like the $N$-body Schr\"odinger equation, is a very active research field in mathematical physics: we refer the reader to the fairly recent monograph~\cite{BenedikterPortaSchlein16} and to the survey article~\cite{Rougerie20} and references therein for an account on the literature.

In this note, we apply this general paradigm to the concrete case of the quantum dynamics of $N$ fermionic particles subject to a uniform external \emph{magnetic field}. Specifically, we work in the \emph{canonical ensemble}, namely we consider the number of particles as fixed (but possibly arbitrarily large). We illustrate how the application of a variational approach known as the \emph{Dirac--Frenkel principle} allows to derive the Hartree--Fock equations as the effective description of such systems, see Theorem~\ref{thm:HF}. These equations provide a mean-field description of the interactions among the fermions, and the solution to the Hartree--Fock equations gives an ``uncorrelated'' approximation to the many-body wave function, whose form is that of a \emph{Slater determinant}: this is the form of the simplest possible wave functions obeying fermionic statistics. In recent times, these effective descriptions of the quantum dynamics in magnetic fields have themselves proved to be useful starting points for considerations e.g.\  in density functional theory and other similar limits~\cite{GontierLahbabiMaichine23, Perice24, PericeRougerie24}, thus motivating the present attempt to give an account on their derivation. We also quantify the effectiveness of the approximation by providing a norm bound for the difference of the solution to the $N$-body Schr\"odinger equation and the outcome of the Dirac--Frenkel variational principle, which is illustrated in Theorem~\ref{thm:Estimate}. We comment on how this bound justifies the nonlinear approximation for small times and for weak interactions, and discuss also the dependence of this bound on the number of particles, in relation to a coupled mean-field and semiclassical scaling. For the sake of a self-contained and introductory presentation, we include a quick overview on the one-particle description of charged quantum particles in a magnetic field through the Landau Hamiltonian (Section~\ref{sec:Landau}), and on generalites on $N$-body quantum systems (Section~\ref{sec:N}). Our exposition shows in particular that translation invariance, which is explicitly broken by any choice of the magnetic vector potential and invalidates the use of the Fourier expansion, is irrelevant for the derivation of the Hartree--Fock equations: indeed, in our presentation, one could replace the eigenvectors of the Landau operator with the plane waves which diagonalize the one-particle free Hamiltonian in the box, and recover in this way the non-magnetic case. While we don't claim any originality in the results described in this note, we hope the reader may still find some usefulness and insights in our presentation of this concrete application of the Dirac--Frenkel principle to quantum dynamics.


\section{One-particle picture: the Landau Hamiltonian} \label{sec:Landau}

We consider a charged quantum particle confined to two dimensions by the presence of a uniform external magnetic field perpendicular to the plane~\cite[Chapter~8]{Thaller00}. The two-dimensional sample could be further confined to a box $\Lambda := [-L_1/2,L_1/2] \times [-L_2/2,L_2/2]$, $L_1, L_2 > 0$, with appropriate boundary conditions to be discussed later, or be infinite (and we will treat both situations in this Section). We choose coordinates in configuration space so that the magnetic field is of the form $\vec{B} = (0,0,B)$ with $q B > 0$, where $q$ is the electric charge of the fermion. If $\vec{A}$ denotes a magnetic potential for the magnetic field $\vec{B}$, i.e.%
\footnote{With a slight abuse of notation, we identify the three-dimensional magnetic vector potential $(A_1, A_2, 0)$ with $\vec{A} = (A_1, A_2) \in \R^2$, and we denote $\vec{x} \wedge \vec{y} := x_1 y_2 - x_2 y_1$ for $\vec{x}=(x_1,x_2)$ and $\vec{y}=(y_1,y_2)$ in~$\R^2$.} %
$\operatorname{curl}\, \vec{A} \equiv (0,0,\vec{\nabla} \wedge \vec{A}) = \vec{B}$, then the one-particle \emph{Landau Hamiltonian} reads
\begin{equation} \label{eqn:Landau1}
H_1 = \frac{1}{2m} \, \left(\vec{p}_{\vec{A}}\right)^2 \equiv \frac{1}{2m} \left( \vec{p} - \frac{q}{c} \vec{A} \right)^2,
\end{equation}
where $m$ is the mass of the particle, $\vec{p} = - \iu \hbar \vec{\nabla}$ is the 2D momentum and $c$ the speed of light. 

Even though we will not make it explicit in the notation, the expression for $H_1$ does depend on the choice of the vector potential. It is well known however that the Landau Hamiltonian displays a \emph{magnetic gauge covariance}, namely that if $\gamma$ is a (smooth) function on $\R^2$ and one sets
\[ \vec{A}' := \vec{A} + \vec{\nabla} \gamma\,, \]
then
\begin{equation} \label{eqn:Hgauge}
H_1' \equiv \frac{1}{2m} \left( \vec{p} - \frac{q}{c} \vec{A}' \right)^2 = \eu^{\iu q \gamma/\hbar c} \left( \frac{1}{2m} \left( \vec{p} - \frac{q}{c} \vec{A} \right)^2 \right) \eu^{-\iu q \gamma/\hbar c} = \eu^{\iu q \gamma/\hbar c} H_1 \eu^{-\iu q \gamma/\hbar c}.
\end{equation}

\subsection{From Landau to the harmonic oscillator}

Observe first of all that the two components of the magnetic momentum $\vec{p}_{\vec{A}} = - \iu \hbar \vec{\nabla} - (q/c) \vec{A}$ do not commute: indeed
\begin{equation} \label{eqn:MagneticComm}
\left[ p_{\vec{A},1}, p_{\vec{A},2} \right] = - \iu \, \hbar \, \frac{q}{c} \vec{\nabla} \wedge \vec{A} = \iu \, \hbar \, \frac{q B}{c} \, \Id \equiv \iu \, \hbar^2 b \, \Id
\end{equation}
where we set, also for future convenience,
\begin{equation} \label{eqn:Dimensionless}
b := \frac{qB}{\hbar c} >0.
\end{equation}
We use this observation to draw a connection between the Landau Hamiltonian and a one-dimensional quantum harmonic oscillator. Define \emph{ladder operators}
\begin{equation} \label{eqn:ladder}
\fa := \iu \, \sqrt{\frac{1}{2 \hbar^2 b}} \left( p_{\vec{A},1} + \iu \, p_{\vec{A},2} \right), \quad \fa^\dagger := -\iu \, \sqrt{\frac{1}{2 \hbar^2 b}} \left( p_{\vec{A},1} - \iu\, p_{\vec{A},2} \right).
\end{equation}
These operators satisfy canonical commutation relation, $[\fa, \fa^\dagger] = \Id$, and therefore $\fa^\dagger \fa$ is a number operator, meaning that $\sigma(\fa^\dagger \fa) = \N$. Moreover, with $\omega\sub{c} := \hbar b / m = q B / m c$ denoting the cyclotron frequency, an immediate check yields
\begin{equation} \label{eqn:H0_from_ladder}
H_1 = \hbar \, \omega\sub{c} \, \left( \fa^\dagger \fa + \frac{1}{2}\, \Id \right) .
\end{equation}
From this, we can conclude at once that the spectrum of $H_1$ consists of the \emph{Landau levels}
\begin{equation} \label{eqn:LandauLevels}
E_n = \hbar \, \omega\sub{c} \left( n + \frac{1}{2} \right) = \frac{\hbar^2}{2m} \, b \, (2n+1) , \quad n \in \N.
\end{equation}

\subsection{Eigenfunctions for the Landau Hamiltonian: infinite volume} \label{sec:InfVolume}

In order to diagonalize the Landau Hamiltonian, we consider first the infinite-volume case $\vec{x} \in \R^2$. For definiteness, we fix the \emph{Landau gauge}
\[ \vec{A}(x_1, x_2) = \vec{A}\sub{L}(x_1,x_2) := B(0,x_1) \]
hereonafter. Since this vector potential does not depend on $x_2$, the standard kinetic momentum in the second direction is a conserved quantity. We can then define a good quantum number $k_2 \in \R$  by performing a partial Fourier reduction: namely, we look for (generalized) eigenstates of the form $\psi_{k_2}(x_1,x_2) = \eu^{\iu k_2 x_2} \phi_{k_2}(x_1)$. The fiber Hamiltonian $H_1(k_2) := \eu^{\iu k_2 x_2} H_1 \eu^{-\iu k_2 x_2}$ is then the restriction of the Landau Hamiltonian to states with fixed momentum $k_2$, that is, to the $\phi$'s. This fiber Hamiltonian describes a one-dimensional quantum harmonic oscillator of cyclotron frequency $\omega\sub{c}$ centered at $k_2/b$: in the Landau gauge
\[
H_1(k_2) = - \frac{\hbar^2}{2m} \frac{\partial^2}{\partial x_1^2} + \frac{1}{2} m \omega\sub{c}^2 \left( x_1 - \frac{k_2}{b} \right)^2 = \hbar \omega_c \left( \fa^\dagger(k_2) \fa(k_2) + \frac{1}{2}\, \Id \right),
\]
where $\fa^\dagger(k_2)$ and $\fa(k_2)$ are fibered ladder operators
\[ \fa(k_2) := \frac{1}{\sqrt{2 b}} \left( \frac{\partial}{\partial x_1} +  b x_1 - k_2 \right), \quad \fa^\dagger(k_2) := \frac{1}{\sqrt{2 b}} \left( - \frac{\partial}{\partial x_1} + b x_1 - k_2 \right). \]
The spectrum of $H(k_2)$ is then also discrete (and independent of $k_2$), formed by the same Landau levels of~\eqref{eqn:LandauLevels}. The eigenfunctions of $H_1(k_2)$ are then also constructed in the standard way: starting from a state annihilated by $\fa(k_2)$, one obtains excited states by subsequent applications of $\fa^\dagger(k_2)$.

This procedure yields a set of generalized eigenstates for the Landau Hamiltonian $H_1$ in the Landau gauge and in infinite volume given by 
\begin{equation} \label{eqn:InfiniteVolEFs}
\varphi^{\infty}_{n,k_2}(x_1,x_2) := \eu^{\iu k_2 x_2} \phi_{n,k_2}^{\infty}(x_1), \quad \text{with } \phi_{n,k_2}^{\infty}(x_1) := b^{1/4} \, h_n\left(\sqrt{b} \left(x_1 - \frac{k_2}{b}\right)\right),
\end{equation}
where the \emph{Hermite functions} are defined by
\begin{equation} \label{HermiteFncts}
h_n(z) := (\sqrt{\pi} \, 2^{n} \, n!)^{-1/2} \, \eu^{-z^2/2} H_n(z)
\end{equation}
with $H_n$ the $n$-th Hermite polynomial. The choice of the numerical constant in $h_n(z)$ is such that, for fixed $k_2 \in \R$, the function $\phi^{\infty}_{n,k_2}$ is normalized in $L^2(\R,\di x_1)$.

\subsection{Magnetic translations}

A similar analysis can be performed in the finite-volume case $\vec{x} \in \Lambda$. We first need to impose (self-adjoint) boundary conditions. To do so, we consider \emph{magnetic translations}. These are generated by (dual) magnetic momenta $\vec{p}_{\vec{A}\,^\vee} = - \iu \hbar \vec{\nabla} - (q/c) \vec{A}\,^\vee$, for some linear vector potential $\vec{A}\,^\vee$, via 
\[ T_{\vec{a}}^{\text{m}} := \eu^{\iu a_1 \, p_{\vec{A}^\vee,1} / \hbar} \, \eu^{\iu a_2 \, p_{\vec{A}^\vee,2} / \hbar}, \quad \vec{a}=(a_1,a_2) \in \R^2. \]
Magnetic translations are required first of all to commute with the Hamiltonian. This can be achieved if
\begin{equation} \label{eqn:DualGauge}
\left[p_{\vec{A},i}, p_{\vec{A}\,^\vee,j} \right] = - \iu \hbar \frac{q}{c} \left( \partial_j A_i - \partial_i A_j^\vee\right) \overset{!}{=} 0 \quad \text{for all } i,j \in \set{1,2}
\end{equation}
since then $[\vec{p}_{\vec{A}}, T_{\vec{a}}^{\text{m}}]=0$ and  $[H_1, T_{\vec{a}}^{\text{m}} ]= [\vec{p}_{\vec{A}}\,^2 / 2m , T_{\vec{a}}^{\text{m}} ]= 0$ as well. Notice that from the above equality it follows that $\vec{\nabla} \wedge \vec{A}\,^\vee = - \vec{\nabla} \wedge \vec{A} = - B$, and hence (compare~\eqref{eqn:MagneticComm})
\begin{equation} \label{eqn:MagneticComm-}
\left[ p_{\vec{A}\,^\vee,1}, p_{\vec{A}\,^\vee,2} \right] = - \iu \, \hbar^2 b \, \Id.
\end{equation}
By using that $\vec{\nabla} \wedge \vec{A} = \partial_1 A_2 - \partial_2 A_1 = B$ is constant, one can solve~\eqref{eqn:DualGauge} and obtain
\begin{equation} \label{eqn:Avee}
\vec{A}\,^\vee(x_1,x_2) = \vec{A}(x_1,x_2) + B \begin{pmatrix} x_2 \\ - x_1 \end{pmatrix}.
\end{equation}

We can then choose \emph{magnetic-periodic boundary conditions} for the Landau Hamiltonian on the box $\Lambda$, namely
\begin{equation} \label{eqn:MPBC}
\begin{gathered}
\varphi(-L_1/2,x_2) =  \left(T_{(L_1,0)}^{\text{m}} \varphi\right)(-L_1/2,x_2), \\ \varphi(x_1,-L_2/2) =  \left(T_{(0,L_2)}^{\text{m}} \varphi\right)(x_1,-L_2/2),
\end{gathered} \qquad\qquad (x_1,x_2) \in \Lambda.
\end{equation}
In order for the two conditions above to be consistent with each other, the two magnetic translations $T_{(L_1,0)}^{\text{m}}$ and $T_{(0,L_2)}^{\text{m}}$ should commute. This is not the case, due to the noncommutativity of the components of the magnetic momenta. To see this more explicitly, we use the Baker--Campbell--Hausdorff formula to elaborate the expression for the magnetic translations. First of all, notice that the commutator in~\eqref{eqn:MagneticComm} is central, that is, it commutes with both $p_{\vec{A}\,^\vee,1}$ and $p_{\vec{A}\,^\vee,2}$. The Baker--Campbell--Hausdorff formula then reduces to
\[
T_{(L_1,0)}^{\text{m}} T_{(0,L_2)}^{\text{m}}  = \eu^{\iu L_1 \, p_{\vec{A}\,^\vee,1} / \hbar} \, \eu^{\iu L_2 \, p_{\vec{A}\,^\vee,2} / \hbar}
= \eu^{\iu b L_1 L_2/2} \eu^{\iu \vec{L} \cdot \vec{p}_{\vec{A}\,^\vee}/\hbar}
\]
where $\vec{L}=(L_1,L_2)$. A similar computation yields
\[ T_{(L_2,0)}^{\text{m}} T_{(0,L_1)}^{\text{m}}  = \eu^{-\iu b L_1 L_2/2} \eu^{\iu \vec{L} \cdot \vec{p}_{\vec{A}\,^\vee}/\hbar} \]
and hence
\[ T_{(L_1,0)}^{\text{m}} T_{(0,L_2)}^{\text{m}} = \eu^{\iu b L_1 L_2} \, T_{(L_2,0)}^{\text{m}} T_{(0,L_1)}^{\text{m}}. \]
In order for magnetic-periodic boundary conditions to be consistent, then, we need to impose the following \emph{quantization condition}:
\begin{equation} \label{eqn:FluxQuantum}
b L_1 L_2 \overset{!}{=} 2 \pi M, \quad M \in \N.
\end{equation}
Notice that the above equality means that the magnetic flux through the sample $\Lambda$ is an integer multiple of the magnetic flux quantum $\Phi_0 := hc/q$:
\[ \Phi_B := \int_{\Lambda} \di \vec{x} \, \vec{\nabla} \wedge \vec{A}(\vec{x}) = B L_1 L_2 \overset{!}{=} 2 \pi M \frac{\hbar c}{q} = M \Phi_0, \quad M \in \N. \]
Under this condition, moreover, the magnetic translations form a unitary representation of the lattice
\[ \Gamma := \set{ \vec{\ell} = (\ell_1 L_1, \ell_2 L_2) : \ell_1, \ell_2 \in \Z } \simeq \Z^2, \]
namely $T\su{m} : \Gamma \to \mathcal{U}(L^2(\Lambda))$, $\vec{\ell} \mapsto T_{\vec{\ell}}\su{m}$, satisfies $T_{\vec{0}}\su{m} = \Id$ and $T_{\vec{\ell}}\su{m} T_{\vec{\ell}'}\su{m} = T_{\vec{\ell}+\vec{\ell}'}\su{m} = T_{\vec{\ell}'}\su{m} T_{\vec{\ell}}\su{m}$.

Concretely, if we fix once again the Landau gauge $\vec{A}\sub{L}(x_1, x_2) = B(0,x_1)$, then $\vec{A}\,^\vee\sub{L}(x_1,x_2) = B(x_2,0)$ from~\eqref{eqn:Avee} and
\begin{equation} \label{eqn:MagTransl}
T_{\vec{a}}^{\text{m}} = \eu^{\iu a_1 (p_1 - qBx_2/c)/\hbar} \, \eu^{\iu a_2 p_2} = \eu^{-\iu \, b \, a_1 \, x_2} \, T_{\vec{a}},
\end{equation}
where $T_{\vec{a}} := \eu^{\iu \vec{a} \cdot \vec{p}/\hbar}$ are the usual translations, acting as
$(T_{\vec{a}} \varphi)(\vec{x}) = \varphi(\vec{x}+\vec{a})$. The magnetic-periodic boundary conditions then read in this gauge
\begin{equation} \label{eqn:MPBC-Landau}
\begin{gathered}
\varphi(-L_1/2,x_2) = \eu^{-\iu \, b \, L_1 \, x_2} \, \varphi(L_1/2,x_2) = \eu^{-\iu \, 2 \pi M \, x_2/L_2} \, \varphi(L_1/2,x_2) , \\ 
\varphi(x_1,-L_2/2) = \varphi(x_1,L_2/2),
\end{gathered} \qquad (x_1,x_2) \in \Lambda.
\end{equation}

\subsection{Eigenfunctions for the Landau Hamiltonian: finite volume}

We now come back to the question of finding eigenfunctions for the Landau Hamiltonian in the Landau gauge satisfing the magnetic-periodic boundary conditions~\eqref{eqn:MPBC-Landau}. One can perform the same analysis from Section~\ref{sec:InfVolume}, the main difference being that the conserved momentum in the second direction $k_2$ now takes values in $(2\pi / L_2) \Z$, and the corresponding Fourier multiplier $\eu^{\iu k_2 x_2}$ can be also $L^2$-normalized by dividing it by a factor of $\sqrt{L_2}$. In order to enforce also the first condition in~\eqref{eqn:MPBC-Landau}, notice that for $\ell_1 \in \Z$ and $k_2 = 2 \pi m / L_2 \in (2 \pi/L_2)\Z$
\begin{equation} \label{eqn:MomentumShift}
\begin{aligned}
\left(T_{(\ell_1 L_1,0)}\su{m} \varphi_{n,2 \pi m/L_2}^{\infty}\right)(\vec{x})& = b^{1/4} \, \eu^{-\iu 2 \pi M \ell_1 x_2 / L_2} \, \eu^{\iu k_2 x_2} h_n \left( \sqrt{b} \left(x_1 + \ell_1 L_1 - \frac{k_2}{b} \right) \right) \\
& = b^{1/4} \, \eu^{\iu 2 \pi (m - M \ell_1) x_2 / L_2} \, h_n \left( \sqrt{b} \left(x_1 - \frac{2 \pi}{L_2} (m - M \ell_1) \right) \right) \\
& =  \varphi_{n,2 \pi (m-M\ell_1)/L_2}^{\infty}(\vec{x})
\end{aligned}
\end{equation}
where we made use of the quantization condition~\eqref{eqn:FluxQuantum} in the second equality. Thus, if we saturate over magnetic translations and define, for $m \in \set{0,\ldots,M-1} \equiv \Z / M\Z$,%
\footnote{Notice that this series is convergent due to the Gaussian decay at infinity of the Hermite functions.}
\begin{equation} \label{eqn:FiniteVolEFs}
\begin{aligned}
\varphi_{n,m}(\vec{x}) & := \frac{1}{\sqrt{L_2}} \sum_{\ell_1 \in \Z}  \left(T_{(\ell_1 L_1,0)}\su{m} \varphi_{n,2 \pi m/L_2}^{\infty}\right)(\vec{x}) = \frac{1}{\sqrt{L_2}} \sum_{\ell_1 \in \Z}  \varphi_{n,2 \pi (m - M \ell_1)/L_2}^{\infty}(\vec{x}) \\
& = b^{1/4} \, \frac{\eu^{\iu 2 \pi m x_2/L_2}}{\sqrt{L_2}} \sum_{\ell_1 \in \Z} \eu^{-\iu 2 \pi M \ell_1 x_2/L_2} \, h_n\left( \sqrt{b} \left(x_1 + \left(\ell_1 - \frac{m}{M}\right) L_1  \right) \right),
\end{aligned}
\end{equation}
we will obtain states which satisfy the magnetic-periodic boundary conditions~\eqref{eqn:MPBC-Landau} and are eigenstates of $H_1$ with energy~$E_n$. These are, moreover, orthonormal. Indeed, notice that if $m \ne m' \in \set{0,\ldots,M-1}$ each state in the sum defining $\varphi_{n,m}$ is orthogonal to each other state in the sum defining $\varphi_{n,m'}$, since they have different momentum $k_2$. It follows that $\scal{\varphi_{n,m}}{\varphi_{n,m'}}_{L^2(\Lambda)} = \delta_{m,m'} \, \norm{\varphi_{n,m}}_{L^2(\Lambda)}^2$. To compute the latter norm squared, notice that again by the orthogonality of the vectors in the sum defining $\varphi_{n,m}$ we can use Parseval's identity to get
\begin{equation} \label{eqn:norm2}
\norm{\varphi_{n,m}}_{L^2(\Lambda)}^2 = \frac{1}{L_2} \sum_{k_2 \in 2 \pi (M \Z + m) / L_2} \norm{\varphi_{n,k_2}^{\infty}}_{L^2(\Lambda)}^2 = \sum_{k_2 \in 2 \pi (M \Z + m) / L_2} \norm{\phi_{0,k_2}^{\infty}}_{L^2([-L_1/2,L_1/2])}^2\,.
\end{equation}
Notice now that in view of \eqref{eqn:FluxQuantum}
\begin{equation} \label{eqn:bL1shift}
\begin{aligned}
\phi_{n,k_2+2\pi M/L_2}^{\infty}(x_1)& = \phi_{n,k_2+bL_1}^{\infty}(x_1) = b^{1/4} \, h_n \left( \sqrt{b} \left( x_1 - \frac{k_2+bL_1}{b} \right) \right) \\
&= b^{1/4} \, h_n \left( \sqrt{b} \left( x_1 - L_1 - \frac{k_2}{b} \right) \right) = \phi_{n,k_2}^{\infty}(x_1-L_1)
\end{aligned}
\end{equation}
so that shifting the momentum $k_2$ by $2\pi M/ L_2$ is equivalent to shifting the argument of the function $\phi_{n,k_2}^{\infty}$ by $-L_1$. The sum in~\eqref{eqn:norm2} becomes then
\[ \norm{\varphi_{n,m}}_{L^2(\Lambda)}^2 = \sum_{\ell_1 \in \Z} \norm{\phi_{0,m}^{\infty}}_{L^2([-L_1/2,L_1/2]-\ell_1 L_1)}^2 = \norm{\phi_{0,m}^{\infty}}_{L^2(\R)}^2 = 1 \]
as wanted. By a similar argument, invoking the completeness in $L^2([-L_2/2,L_2/2],\di x_2)$ of the Fourier multipliers $\set{\eu^{\iu k_2 x_2}/\sqrt{L_2}}_{k_2 \in (2 \pi/L_2) \Z}$ and the completeness of the Hermite functions as an orthonormal set in $L^2(\R,\di x_1)$ \cite{Szego75}, one can conclude that the collection $\set{\varphi_{n,m}}_{n \in \N, \, m \in \Z/M\Z}$ forms an orthonormal basis for $L^2(\Lambda)$.

We collect the previous considerations in the following statement.

\begin{theorem} \label{thm:Landau1}
The operator $H_1$ defined in~\eqref{eqn:Landau1}, acting on $L^2(\Lambda)$ with magnetic-periodic boundary conditions~\eqref{eqn:MPBC}, is self-adjoint on the domain given by the \emph{magnetic Sobolev space}
\[ \cD(H_1) = H^2_{\vec{A}}(\Lambda) := \set{ \psi \in L^2(\Lambda) : H_1 \, \psi \in L^2(\Lambda)}\,. \]
The spectrum of $H_1$ consists of Landau levels 
\[ \sigma(H_1) = \sigma\sub{p}(H_1) = \set{E_n = \frac{\hbar^2}{2m} \, b \, (2n+1) : n \in \N} \]
and moreover
\[ \ker \left( H_1 - E_n \, \Id \right) = \Span \set{\varphi_{n,m} : m \in \set{0,\ldots,M-1} \equiv \Z / M \Z}, \quad n \in \N\,, \]
with $\varphi_{n,m}$ as in~\eqref{eqn:FiniteVolEFs}. In particular, each Landau level has finite degeneracy $M$, with $M \in \N$ as in~\eqref{eqn:FluxQuantum}. Finally, the collection $\set{\varphi_{n,m}}_{n \in \N, \, m \in \Z/M\Z}$ forms an orthonormal eigenbasis of $H_1$ for $L^2(\Lambda)$.
\end{theorem}
\begin{proof}
The only point left to prove is to show that $H_1 = H_1^*$ is self-adjoint. First of all, $H_1^*$ extends $H_1$ because the latter is symmetric, as can be easily seen by integration by parts: the boundary terms on $\partial \Lambda$ vanish because of magnetic-periodic boundary conditions. To show that the converse is also true, we need to show that $\mathcal{D}(H_1^*) \subset \mathcal{D}(H_1) = H^2_{\vec{A}}(\Lambda)$, where
\[ \mathcal{D}(H_1^*) := \set{f \in L^2(\Lambda) :  \exists \, g \in L^2(\Lambda) :  \scal{H_1\varphi}{f} = \scal{\varphi}{g} \; \forall \, \varphi \in \mathcal{D}(H_1)}\,. \]
Let then $f \in \mathcal{D}(H_1^*) \subset L^2(\Lambda)$: we want to show that also $H_1 f \in L^2(\Lambda)$. Since the set $\set{\varphi_{n,m}}_{n \in \N,\, m \in \Z / M \Z}$ is orthonormal and complete, by Parseval's identity it suffices to show that
\[ \sum_{\substack{n \in \N \\ m \in \Z / M \Z}} \left| \scal{\varphi_{n,m}}{H_1 f} \right|^2 < + \infty\,. \]
By the symmetry of $H_1$ and the definition of $\mathcal{D}(H_1^*)$, we have that
\[ \scal{\varphi_{n,m}}{H_1 f} = \scal{H_1\varphi_{n,m}}{f} = \scal{\varphi_{n,m}}{g} \]
for some $g \in L^2(\Lambda)$, independent of $n \in \N$ and $m \in \Z /M\Z$, since all $\varphi_{n,m}$'s are in the domain of $H_1$. But then
\[ \sum_{\substack{n \in \N \\ m \in \Z / M \Z}} \left| \scal{\varphi_{n,m}}{H_1 f} \right|^2 = \sum_{\substack{n \in \N \\ m \in \Z / M \Z}} \left| \scal{\varphi_{n,m}}{g} \right|^2 = \norm{g}_{L^2(\Lambda)}^2 < + \infty \]
which concludes the proof. 
\end{proof}

\section{$N$-particle picture: non-interacting ground state} \label{sec:N}

The dynamics of a single quantum particle in an external uniform magnetic field is described by the Landau Hamiltonian, whose properties were illustrated in the previous Section. When a quantum system is composed by a number $N \ge 2$ of particles, then the effects of the interactions between such particles should be included in the model. We recall in this Section a few features of many-body quantum systems which are relevant for our discussion: the reader is referred e.g.\ to~\cite{LiebSeiringer09} for a more detailed description.

\subsection{Generalities on $N$-body quantum systems}

First of all, the wave function of the collective system depends on the $N$ positions $\vec{x}_1, \ldots, \vec{x}_N$ of the particles, and are thus elements of the Hilbert space 
\[ L^2(\Lambda^{N}) \simeq \bigotimes_{i=1}^{N} L^2(\Lambda) \equiv L^2(\Lambda) \otimes \stackrel{N \text{ times}}{\cdots} \otimes L^2(\Lambda) \]
if $\Lambda$ is the configuration space for the single particle (e.g.\ the 2-dimensional box). Actually, wave functions should be described as vector-valued, i.e.\ elements of $L^2(\Lambda^{N}) \otimes \C^S$, to take into account the \emph{spin} of the quantum particles they model. Many-body systems reveal however a new feature, peculiar to quantum particles, in that the particles' spin influences their \emph{statistics}: this feature is modelled by a phase which is picked up by the wave function, when the positions of two particles are exchanged. Particularly relevant are the cases in which this phase is $+1$ --- i.e.\ \emph{totally symmetric} wave functions, describing \emph{bosonic particles} or \emph{bosons}, for simplicity --- and the one in which this phase is $-1$ --- i.e.\ \emph{totally antisymmetric} wave functions, describing \emph{fermionic particles} or \emph{fermions}, for simplicity. In the following, we will be interested in fermionic wave functions, and make the simplified, ``academic'' assumption of treating \emph{spinless fermions}: this means that we ignore the vectorial spin degrees of freedom of the particles, and model them through scalar, antisymmetric wave functions in the subspace
\[ \cH_N = \bigwedge_{i = 1}^{N} L^2(\Lambda) := L^2(\Lambda) \, \wedge \stackrel{N \text{ times}}{\cdots} \wedge \, L^2(\Lambda) \subset L^2(\Lambda^{N})\,. \]
Thus, the main feature of the particles' statistics which is retained by this description is the Pauli exclusion principle, which dictates that two particles cannot be in the same one-particle quantum state $\psi$, as guaranteed by the antisymmetry of the wedge product: $\psi \wedge \psi = 0$.

The simplest wave functions in $\cH_N$ are \emph{Slater determinants}: given $N$ one-particle quantum states $\psi_1, \ldots, \psi_N \in L^2(\Lambda)$ (i.e.\ an orthonormal set of square-integrable functions), we define
\begin{equation} \label{eqn:wedgedef}
\Psi\sub{Slater}(\vec{x}_1, \ldots, \vec{x}_N) = \psi_1 \wedge \cdots \wedge \psi_N(\vec{x}_1, \ldots, \vec{x}_N) := \frac{1}{\sqrt{N!}} \, \det \left[ \psi_i\left(\vec{x}_j\right) \right]_{1 \le i, j \le N}\,. 
\end{equation}
The factor $1/\sqrt{N!}$, which enters in the definition of the wedge product, ensures that
\begin{equation} \label{eqn:ScalarSlater}
\scal{\Psi\sub{Slater}}{\Phi\sub{Slater}}_{\cH_N} = \scal{\psi_1 \wedge \cdots \wedge \psi_N}{\varphi_1 \wedge \cdots \wedge \varphi_N}_{\cH_N} = \det \left[ \scal{\psi_i}{\varphi_j}_{L^2(\Lambda)} \right]_{1 \le i, j \le N}.
\end{equation}
In particular, $\Psi\sub{Slater}$ is normalized in $L^2(\Lambda^N)$ if $\set{\psi_1, \ldots, \psi_N}$ is an orthonormal set in $L^2(\Lambda)$. By definition of the subspace $\cH_N \subset L^2(\Lambda^N)$, Slater determinants span $\cH_N$, and in particular the Slater determinants
\begin{equation} \label{eqn:Slaternm}
\set{\varphi_{n_1,m_1} \wedge \cdots \wedge \varphi_{n_N,m_N} : (n_1,m_1) \ne \cdots \ne (n_N,m_N) \in \N \times \Z / M \Z}
\end{equation}
form an orthonormal basis of $\cH_N$, in view of Theorem~\ref{thm:Landau1} and of~\eqref{eqn:ScalarSlater}.

\subsection{$N$-body quantum Hamiltonian} \label{sec:HN}

Having established the basic notation regarding the kinematics of many-body quantum systems, we are ready to describe their dynamics, i.e.\ the typical quantum Hamiltonian for such systems.  It is customary to neglect, in first approximation, any interaction involving a higher number of particles (like three-body interactions and so on), and therefore to assume that particles interact at most pair-wise. One considers then Hamiltonians on $\cH_N$ of the form
\begin{equation} \label{eqn:HN}
H_N = {H}\sub{ni} + H\sub{int}
\end{equation}
where:
\begin{itemize}
 \item ${H}\sub{ni}$ is the non-interacting part of the Hamiltonian, namely an operator of the form
 \[ {H}\sub{ni} = \sum_{i=1}^{N} \Id_{L^2(\Lambda, \di \vec{x}_1)} \otimes \cdots \otimes \Id_{L^2(\Lambda, \di \vec{x}_{i-1})} \otimes H_1^{(i)} \otimes \Id_{L^2(\Lambda, \di \vec{x}_{i+1})} \otimes \cdots \otimes \Id_{L^2(\Lambda, \di \vec{x}_{N})}\,; \]
 the operator ${H}\sub{ni}$ is thus a sum of operators, each of which acts non-trivially only on the Hilbert space $L^2(\Lambda, \di \vec{x}_i)$ of the $i$-th particle through the operator $H_1^{(i)}$, $i \in \set{1,\ldots,N}$;
 \item $H\sub{int}$ is the interacting part of the Hamiltonian, which accounts for many-body interaction and is of the form
 \begin{equation} \label{eqn:hatV}
 H\sub{int} = \sum_{1 \le i < j \le N} V_{ij} \equiv \sum_{1 \le i < j \le N} V(\vec{x}_i;\vec{x}_j)
 \end{equation}
 where $V_{ij}$ is the multiplication operator times the  \emph{two-body potential} $V(\vec{x}_i;\vec{x}_j)$, depending non-trivially only on the positions of the $i$-th and $j$-th particle, $i,j \in \set{1, \ldots, N}$, which accounts for interactions between this pair of particles. In order to preserve \emph{indistinguishability} of the $N$ quantum particles, it is assumed that the function $V(\cdot, \cdot)$ is left invariant by the exchange of its two entries, namely
 \[ V(\vec{x}; \vec{y}) = V(\vec{y};\vec{x}) \quad \text{for all } \vec{x}, \vec{y} \in \Lambda\,. \]
\end{itemize}

The natural choice for us will be to posit that the non-interacting part of the Hamiltonian $H_N$ is modelled after the one-particle Landau Hamiltonian described in the last Section, namely that
\[ H_1^{(i)} \equiv H_1 \quad \text{for all} \quad i \in \set{1,\ldots,N}, \]
and that $H_1$ is endowed with magnetic-periodic boundary conditions on $\Lambda$. Moreover, we assume that the two-body potential $V$ is a bounded function on $\Lambda \times \Lambda$, so that $H\sub{int}$ is a bounded operator on $\cH_N$ with
\begin{equation} \label{eqn:normhatV}
\norm{H\sub{int}}_{\cB(\cH_N)} \le \binom{N}{2} \, \norm{V}_{L^\infty(\Lambda \times \Lambda)} \,.
\end{equation}
This is a simplifying assumption, and much larger classes of singular potentials (including e.g.\ those modeling Coulomb interactions) can be treated with the appropriate mathematical care: we refer the interested reader to~\cite{LiebSeiringer09, BenedikterPortaSchlein16} and references therein.

\begin{proposition} \label{prop:HNsa}
The Hamiltonian $H_N = {H}\sub{ni} + H\sub{int}$ is self-adjoint on the domain
\[ \cD({H}\sub{ni}) := \bigwedge_{i=1}^{N} H^2_{\vec{A}}(\Lambda)\,. \]
\end{proposition}
\begin{proof}
Recall that the states~\eqref{eqn:Slaternm} generate $\cH_N$ orthonormally. Using~\eqref{eqn:ScalarSlater} and the definition of ${H}\sub{ni}$, it follows from Theorem~\ref{thm:Landau1} that
\[ \scal{\varphi_{n_1,m_1} \wedge \cdots \wedge \varphi_{n_N,m_N}}{{H}\sub{ni} \, \varphi_{n_1,m_1} \wedge \cdots \wedge \varphi_{n_N,m_N}} = \sum_{i=1}^{N} \scal{\varphi_{n_i,m_i}}{H_1 \, \varphi_{n_i,m_i}} = \sum_{i=1}^{N} E_{n_i}\,. \]
Therefore, the states in~\eqref{eqn:Slaternm} form an orthonormal eigenbasis of ${H}\sub{ni}$ for $\cH_N$. Repeating the argument contained in the proof of Theorem~\ref{thm:Landau1}, we deduce that ${H}\sub{ni}$ is self-adjoint on the domain $\cD({H}\sub{ni}) = \bigwedge_{i=1}^{N} \cD(H_1)$. The Kato--Rellich theorem guarantees that $H_N = {H}\sub{ni} + H\sub{int}$ is then self-adjoint on the same domain, as $H\sub{int}$ is a bounded perturbation.
\end{proof}

\subsection{Non-interacting ground states}

A typical quantity of interest to be determined in many-body quantum systems, especially in connection with the question of stability of matter \cite{LiebSeiringer09}, is the \emph{ground state energy}
\[ E_0^{(N)} := \inf_{\substack{\Psi \in \cH_N \\ \norm{\Psi} = 1}} \scal{\Psi}{H_N \, \Psi}\,. \]
For example, the condition $E_0^{(N)} > -\infty$ is called \emph{stability of the first kind} for the quantum system modelled by $H_N$. States $\Psi \in \cH_N$ realizing the infimum in the definition of $E_0$ are called \emph{ground states} for $H_N$. This topic has drawn a lot of attention: restricting to a selection of mathematical results on properties of the ground state energy for systems of interacting fermions (in three-dimensions and without magnetic fields), in particular to its dependence on the number of particles or more generally on the particle density under various assumptions on the two-body interaction potentials, we refer the reader to the monograph \cite{BenedikterPortaSchlein16} and to the more recent works \cite{BenedikterNamPortaSchleinSeiringer21, ChristiansenHainzlNam23CMP, ChristiansenHainzlNam23, FalconiGiacomelliHainzlPorta21, Giacomelli23JFA, Giacomelli23}.

The computation of the ground state energy and of the corresponding ground states is greatly simplified in the non-interacting case $V \equiv 0$. In our present framework, the complete knowledge of the spectral information on the one-particle Landau Hamiltonian $H_1$, provided by Theorem~\ref{thm:Landau1}, translates in a corresponding description of the spectral properties of ${H}\sub{ni}$ as detailed in Proposition~\ref{prop:HNsa}, and therefore in particular of its ground state properties. The non-interacting ground state is constructed by ``filling'' the Landau levels from the lowest up, according to their multiplicity, with the only constraint being given by the Pauli exclusion symmetry (that is, by the use of wedge products or Slater determinants of the one-body energy states).  We summarize the conclusions of this construction in the following statement.

\begin{theorem} \label{thm:NIGS}
Let $H_N = {H}\sub{ni} = \sum_{i=1}^{N} \Id_{\bigwedge_{j=1}^{i-1} L^2(\Lambda)} \otimes H_1 \otimes \Id_{\bigwedge_{j=i+1}^{N} L^2(\Lambda)}$. Write 
\[ N = (\nu + 1) M + r\,, \quad \nu \in \N, \; r \in \set{0, \ldots, M-1}\,, \]
where $M$ is as in~\eqref{eqn:FluxQuantum}. Then the ground state energy of ${H}\sub{ni}$ is
\[ E_0^{(N)} = M \, \sum_{n=0}^{\nu} E_n + r E_{\nu+1} \,, \]
where the $E_n$'s are the Landau levels~\eqref{eqn:LandauLevels}. Moreover, the space of ground states is
\begin{multline*}
\ker \left({H}\sub{ni} - E_0^{(N)} \, \Id_{\cH_N} \right) \\
= \Span \set{ \left(\bigwedge_{n=0}^{\nu} \bigwedge_{m\in \Z/M\Z} \varphi_{n,m} \right)\wedge \varphi_{\nu+1,m_1} \wedge \ldots \wedge \varphi_{\nu+1,m_r} : m_1 \ne \cdots \ne m_r \in \Z / M \Z } \, .
\end{multline*}
In particular, the ground state energy has a degeneracy equal to
\[ \dim \ker \left({H}\sub{ni} - E_0^{(N)} \, \Id_{\cH_N} \right) = \binom{M}{r}\,. \]
\end{theorem}

\begin{remark} \label{rmk:E0N}
From the explicit expression~\eqref{eqn:LandauLevels} for the Landau levels $E_n$'s, we can write the non-interacting ground state energy $E_0^{(N)}$ from the above statement as
\begin{align*}
E_0^{(N)} & = M \, E_0 \, \left[ \sum_{n=0}^{\nu} (2\,n+1) + r \, (2 \, \nu + 3) \right] = M \, E_0 \, \left[ (\nu+1)^2 + r \, (2 \, \nu + 3) \right] \\
& = M \, E_0 \, \left[ \left(\frac{N-r}{M} \right)^2 + r \left(1 + 2\, \frac{N-r}{M} \right) \right]\,.
\end{align*}
We notice in particular, for future reference, that the dependence of $E_0^{(N)}$ on the number of particles is \emph{quadratic} for large $N$.
\end{remark}

\section{Hartree--Fock effective dynamics for interacting fermions}

When fermions interact, that is, when $V \not\equiv 0$, the description of their dynamics becomes more involved. Even though the Schr\"{o}dinger equation generated by the Hamiltonian $H_N$ is linear, and thefore its solution can be expressed as $\Psi(t) = \eu^{-\iu H_N t/\hbar} \Psi(0)$, the state $\Psi(t)$ at $t\ne0$ is typically difficult to describe (e.g.\ to compute numerically), as correlations between particles are generated by the interacting potential even when the initial state $\Psi(0)$ is minimally correlated (that is, when it is a Slater determinant). One then looks for a simpler, \emph{effective} description of the dynamics of the quantum state. When $\Psi(0)$ is a Slater determinant --- for example one of the non-interacting ground states described in Theorem~\ref{thm:NIGS} --- a possible approach is to try and ``force'' the evolution to stay in the manifold of Slater determinants. While Slater determinants are definitely easier to handle, since one can ``separate'' the orbitals describing the dynamical evolution of the different particles, the result of this approach is that one trades the difficulty of having a many-body wave function $\Psi(t)$ depending on the positions of the $N$ particles all at once with having to describe $N$ one-body states, which turn out to be coupled \emph{nonlinearly} among each other.

We will now describe this method, which is due to Dirac~\cite{Dirac30} and Frenkel~\cite{Frenkel34}, to deduce the effective nonlinear dynamics within the space of Slater determinants from the linear Schr\"{o}dinger dynamics by means of a variational approach. We follow the presentation in~\cite{Lubich08}, to which the reader is referred for further details. As we will detail in the next Section~\ref{sec:Estimate}, our aim is to show that the effective dynamics gives a good approximation of the actual dynamics at least for short times and for weak interactions, as expected from the previous discussion. We will comment also on the dependence of this approximation on the number of particles $N$. It is worth noting that a different application of the same Dirac--Frenkel principle to derive an effective Hartree--Fock dynamics for non-magnetic fermionic particle systems, formulated in terms of density matrices acting on the antisymmetric Fock space (so, in the grand-canonical picture), has been described in~\cite{BenedikterSokSolovej18}.

\subsection{Dirac--Frenkel principle}

In a general setting, the Dirac--Frenkel variational principle to derive an effective equation from Schr\"{o}dinger's can be formulated as follows. Consider a Hamiltonian $H$ that is a self-adjoint and linear operator on a Hilbert space $\cH$. Let $\cM$ be a smooth submanifold of $\cH$, and for $u \in \cM$, let $T_u\cM$ denote the (complex) tangent space to $\cM$ at $u$: by definition, $T_u\cM$ consists of velocity vectors to every differentiable path on $\cM$ passing through $u$. We want to obtain a path $u(t) \in \cM$, with initial condition $u(0) = \Psi(0) \in \cM$, which at least for small $t$ approximates the solution $\Psi(t) = \eu^{-\iu H t/\hbar} \Psi(0)$ to the Schr\"{o}dinger equation. The \emph{Dirac--Frenkel principle} states that $u(t) \in \cM$ should be chosen such that, at each time $t$, the derivative $\dot{u}(t)$ lying on the tangent space $T_{u(t)}\cM$ satisfies the following condition:
\begin{equation} \label{condDirac}
\scal{v}{\iu \,\hbar\, \dot{u}(t) - H\,u(t)} = 0 \quad \text{for all } v \in T_{u(t)}\cM \,.
\end{equation}

The above can be seen as a variational principle: indeed, by taking the real part of the above scalar product, one can realize that~\cite{Lubich08} 
\[ \dot{u}(t) = \mathop{\operatorname{argmin}}_{w \in T_{u(t)}\cM} \, \norm{w - \frac{1}{\iu \,\hbar} H \, u(t)}_{\cH}\,. \]

\begin{proposition} \label{prop:Preservation}
If $u(t)$ satisfies the Dirac--Frenkel principle~\eqref{condDirac}, then the flow $t \mapsto u(t)$ preserves the energy:
\[ \scal{u(t)}{H\,u(t)} \equiv \scal{u(0)}{H\,u(0)} =\scal{\Psi(0)}{H\,\Psi(0)} \quad \text{for all } t\,. \]
If moreover $u(t) \in T_{u(t)} \cM$, then the flow $t \mapsto u(t)$ also preserves the norm:
\[ \norm{u(t)} \equiv \norm{u(0)} = \norm{\Psi(0)} \quad \text{for all } t \,. \]
\end{proposition}
\begin{proof}
As for the preservation of energy, it suffices to compute
\begin{align*}
\frac{\di}{\di t} \scal{u(t)}{H\,u(t)} & = \scal{\dot{u}(t)}{H\,u(t)} + \scal{u(t)}{H\,\dot{u}(t)} = \scal{\dot{u}(t)}{H\,u(t)} + \overline{\scal{\dot{u}(t)}{H\,u(t)}} \\
& = 2 \, \re \scal{\dot{u}(t)}{H\,u(t)} = -2\hbar\, \im \scal{\dot{u}(t)}{\frac{1}{\iu \, \hbar} \,H\,u(t)} \\
& = - 2 \hbar \, \im \norm{\dot{u}(t)}^2_{\cH} = 0\,.
\end{align*}
In the second-to-last equality, we have used~\eqref{condDirac} for $v = \dot{u}(t)$. 

Concerning instead the preservation of the norm, we can similarly compute
\begin{align*}
\frac{\di}{\di t} \scal{u(t)}{u(t)} & = \scal{\dot{u}(t)}{u(t)} + \scal{u(t)}{\dot{u}(t)} = \scal{u(t)}{\dot{u}(t)} + \overline{\scal{u(t)}{\dot{u}(t)}} \\
& = 2 \, \re \scal{u(t)}{\dot{u}(t)} = \frac{2}{\hbar}\, \im \scal{u(t)}{\iu \, \hbar\, \dot{u}(t)} \\
& = \frac{2}{\hbar} \, \im \scal{u(t)}{H\,u(t)} = 0
\end{align*}
due to the self-adjointness of $H$. Again, in the second-to-last equality, we have used~\eqref{condDirac} for $v = u(t) \in T_{u(t)} \cM$.
\end{proof}

\subsection{Hartree--Fock equations from the Dirac--Frenkel principle}

We now apply the Dirac--Frenkel variational principle to the $N$-body Hamiltonian $H_N$ from~\eqref{eqn:HN} on $\cH_N$. We choose the manifold of Slater determinants for the restricted dynamics:
\begin{equation} \label{varietàHF}
\cM := \set{u = a \, \varphi_1 \wedge \cdots \wedge \varphi_N : a \in \mathbb{C}, \, \varphi_i \in L^2(\Lambda),\, \scal{\varphi_i}{\varphi_j} =  \delta_{i,j} \text{ for } i,j \in \set{1, \ldots,N}}\,.
\end{equation}
In the above, we have parametrized Slater determinants with orthonormal orbitals: in particular, in this way, we have that $\norm{u} = |a|$. Notice that $\cM$ contains rays, that is, if $u$ is in $\cM$ then $s \, u$ is in $\cM$ for all $s \in \R$; therefore, $u = (\di/\di s) (s \, u) \big|_{s=0} \in T_u \cM$ for all $u \in \cM$. Then, Proposition~\ref{prop:Preservation} applies, and the solution $u(t) \in \cM$ of the Dirac--Frenkel variational principle~\eqref{condDirac} will have constant norm for all $t$. We will take $u(0) = \Psi(0) \in \cM \subset \cH_N$ of unit norm, as is natural for the initial datum of the Schr\"{o}dinger equation; therefore in particular 
\begin{equation} \label{eqn:|a|}
|a(t)| = \norm{u(t)} \equiv 1 \quad \text{for all } t\,. 
\end{equation}

From~\eqref{varietàHF} it appears that a natural choice to parametrize tangent vectors $\dot{u} \in T_u\cM$ is
\begin{equation} \label{equdot}
\dot{u} = \dot{a} \, \varphi_1 \wedge \cdots \wedge \varphi_N + a \, \dot {\varphi}_1 \wedge \varphi_2 \wedge \cdots \wedge \varphi_N + \cdots + a \, \varphi_1 \wedge \cdots \wedge \varphi_{N-1} \wedge \dot{\varphi}_N \,,
\end{equation}
with $\dot{a} \in \mathbb{C}$ and $\dot{\varphi}_i \in L^2(\Lambda)$, $i \in \set{1,\ldots,N}$. If $u(t) \in \cM$ is the solution to the Dirac--Frenkel variational principle, this parametrization for $\dot{u}(t) \in T_{u(t)} \cM$ can be chosen so that it satisfies further constraints.

\begin{proposition} \label{prop:GaugeFreedom}
Let $t \mapsto u(t) \in \cM$ be the solution to~\eqref{condDirac}. Then $\dot{u}(t) \equiv \di u(t)/ \di t \in T_{u(t)} \cM$ can be chosen as in~\eqref{equdot} with
\begin{equation} \label{eqn:GaugeFreedom}
\scal{\varphi_i(t)}{\dot{\varphi}_j(t)} = 0 \quad \text{for all } i,j \in \set{1,\ldots,N} \text{ and all } t\,.
\end{equation}
\end{proposition}
\begin{proof}
It is clear that, if $U \in \mathrm{U}(N)$ is an $N\times N$ unitary matrix, then 
\[ a \, \varphi_1 \wedge \cdots \wedge \varphi_N = a' \, \varphi_1' \wedge \cdots \wedge \varphi_N' \]
where
\[ a' = \frac{a}{\det U} \quad \text{and} \quad \varphi_i' = \sum_{j=1}^{n} \varphi_j \, U_{ji}\,. \]
This is because linear combination of orthonormal vectors with coefficients from a unitary matrix yield still orthonormal vectors, and $\scal{\varphi_1 \wedge \cdots \wedge \varphi_N}{\varphi_1' \wedge \cdots \wedge \varphi_N'} = \det U$ by the properties of the scalar product of Slater determinants of orthonormal orbitals. We refer to this as a \emph{gauge freedom} in the representation of $u = a \, \varphi_1 \wedge \cdots \wedge \varphi_N \in \cM$ with $|a|=\norm{u}$ and $\varphi_1, \ldots, \varphi_N$ orthonormal. Similarly, we have a gauge freedom in the parametrization~\eqref{equdot} of a tangent vector $\dot{u} \in T_u \cM$: indeed, writing
\[ \dot{\varphi}_j = \sum_{i=1}^{N} \scal{\varphi_i}{\dot{\varphi}_j} \, \varphi_i + \dot{\varphi}_j', \quad \text{with } \scal{\varphi_i}{\dot{\varphi}_j'} = 0 \text{ for all } i, j \in \set{1,\ldots, N}\,, \]
we can have
\begin{multline*}
\dot{a} \, \varphi_1 \wedge \cdots \wedge \varphi_N + a \, \sum_{j=1}^{N} \varphi_1 \wedge \cdots \wedge \dot {\varphi}_j \wedge \cdots \wedge \varphi_N \\ 
= \dot{a}'  \, \varphi_1 \wedge \cdots \wedge \varphi_N + a \, \sum_{j=1}^{N} \varphi_1 \wedge \cdots \wedge \dot {\varphi}_j' \wedge \cdots \wedge \varphi_N
\end{multline*}
by setting
\[ \dot{a}' = \dot{a} - \sum_{j=1}^{N} \scal{\varphi_j}{\dot{\varphi}_j}\,. \]
Using this reparametrization at $t=0$, we will therefore assume without loss of generality that 
\begin{equation} \label{eqn:orthodot}
\scal{\varphi_i(0)}{\dot{\varphi}_j(0)} = 0 \quad \text{for all } i,j \in \set{1, \ldots, N}\,.
\end{equation}

We now want to exploit this gauge freedom to guarantee that the conditions in the statement are satisfied. Consider then the solution $u(t) \in \cM$ of the Dirac--Frenkel principle. First of all, let us notice that if the orbitals $\varphi_i(t)$ which describe $u(t)$ have to be orthonormal at all times, then we should require that for all $i, j \in \set{1,\ldots,N}$
\[ 0 = \frac{\di}{\di t} \scal{\varphi_i(t)}{\varphi_j(t)} = \scal{\varphi_i(t)}{\dot{\varphi}_j(t)} +\scal{\dot{\varphi}_i(t)}{\varphi_j(t)} = \scal{\varphi_i(t)}{\dot{\varphi}_j(t)} + \overline{\scal{\varphi_j(t)}{\dot{\varphi}_i(t)}}\,. \]
The above condition can be recast by saying that the matrix
\[ B(t) = \big[ B(t)_{ij} \big]_{1 \le i,j \le N}\,, \quad B(t)_{ij} := \scal{\varphi_i(t)}{\dot{\varphi}_j(t)}\,, \]
is skew-adjoint, $B(t)^* = - B(t)$, and satisfies $B(0) = 0$ by~\eqref{eqn:orthodot}. Define now for $i \in \set{1,\ldots,N}$
\[ \varphi_i'(t) := \sum_{j=1}^{n} \varphi_j(t) \, U(t)_{ji}\,, \quad U(t) = \big[ U(t)_{ij} \big]_{1 \le i,j \le N} \in \mathrm{U}(N) \,. \]
We show that there exists a choice of $U(t) \in \mathrm{U}(N)$ such that the orthonormal vectors $\set{\varphi_1'(t), \ldots, \varphi_N'(t)}$ satisfy the condition in~\eqref{eqn:GaugeFreedom} at all $t$. Indeed, it is easy to compute that
\[ B'(t)_{ij} := \scal{\varphi_i'(t)}{\dot{\varphi}_j'(t)} \quad \text{is such that} \quad B'(t) = U(t)^* \, B(t) \, U(t) + U(t)^* \, \dot{U}(t)\,. \]
Therefore, $U(t)$ should be chosen as the solution to the Cauchy problem
\[ \begin{cases} \dot{U}(t) = - B(t) \, U(t) \,, \\ U(0) = \Id_N\,, \end{cases} \]
which has a unique unitary solution in view of the skew-adjointness of the generator $-B(t)$.
\end{proof}

Having imposed the gauge condition~\eqref{eqn:GaugeFreedom}, let us now derive the equations of motion for the parameters appearing in~\eqref{equdot}, namely the so-called \emph{Hartree--Fock equations}: these dictate the effective non-linear dynamics of an initially factorized state, in the form of a Slater determinant, which as mentioned above could be for example a ground state for the non-interacting Hamiltonian as presented in Theorem~\ref{thm:NIGS}.

\begin{theorem} \label{thm:HF}
Let $u(t) \in \cM$ be the solution to~\eqref{condDirac}, with 
\begin{equation} \label{eqn:accaenne}
H_N = {H}\sub{ni} + H\sub{int} = \sum_{1 \le i \le N} {H}^{(i)} + \sum_{1 \le i < j \le N} V_{ij}\,, \quad {H}^{(i)} := \Id_{\bigwedge_{j=1}^{i-1} L^2(\Lambda)} \wedge H_1 \wedge \Id_{\bigwedge_{j=i+1}^{N} L^2(\Lambda)}
\end{equation}
of the form described in Section~\ref{sec:HN} --- in particular, $\Lambda = [-L_1/2, L_1/2] \times [-L_2/2, L_2/2]$ is a box and $V \in L^\infty(\Lambda \times \Lambda)$. Then 
\[ u(t) = a(t) \, \varphi_1(t) \wedge \cdots \wedge \varphi_N(t) \]
where $a(t) = \eu^{-\iu \, \cE_0^{(N)} \, t / \hbar}\, a(0)$ satisfies the differential equation
\begin{equation} \label{eqn:dota}
\iu \, \hbar \, \dot{a}(t) = \cE_0^{(N)} \, a(t) \quad \text{for} \quad \cE_0^{(N)} := \scal{u(0)}{H_N \, u(0)} = \scal{\Psi(0)}{H_N \, \Psi(0)}
\end{equation}
and the orthonormal orbitals $\set{\varphi_1(t), \ldots, \varphi_N(t)}$ satisfy the \emph{Hartree--Fock equations}
\begin{equation} \label{eqn:HF}
\iu \, \hbar \, \dot{\varphi}_\ell(t) = H_1 \, \varphi_\ell(t) + K_\ell(t) \, \varphi_\ell(t) - \sum_{\ell' \ne \ell} X_{\ell,\ell'}(t) \, \varphi_{\ell'}(t)\,, \quad \ell \in \set{1, \ldots, N}\,,
\end{equation}
where the \emph{Hartree--Fock potential} $K_\ell$ and the \emph{exchange potentials} $X_{\ell,n}$ are defined as follows:
\begin{gather}
K_\ell(t, \vec{x}) := \sum_{\ell' \ne \ell} \int_{\Lambda} \di \vec{y} \; V(\vec{x};\vec{y}) \, \left| \varphi_{\ell'}(t, \vec{y}) \right|^2\,, \ell \in \set{1, \ldots, N}\,, \label{eqn:KHF} \\
X_{\ell,\ell'}(t, \vec{x}) := \int_{\Lambda} \di \vec{y} \, V(\vec{x};\vec{y}) \; \overline{\varphi_{\ell'}(t,\vec{y})} \, \varphi_\ell(t,\vec{y})\,, \quad \ell,\ell' \in \set{1, \ldots, N}\,. \label{eqn:XHF}
\end{gather}

The Hartree--Fock equations constitute a system of (at least locally) well-posed nonlinear partial differential equations, that is, there exist a time interval $0 \le t \le \bar{t}$ for which a solution $\varphi_\ell \in C^1([0,\bar{t}], L^2(\Lambda)) \cap C([0, \bar{t}], H_{\vec{A}}^2(\Lambda))$ of the Hartree--Fock equations exists. In particular, orthonormality of the $\varphi_\ell(t)$'s solutions to~\eqref{eqn:HF} is preserved throughout the entire time interval.
\end{theorem}
\begin{proof}
Let us rewrite the parametrization~\eqref{equdot} for the generic vector $v \in T_{u(t)} \cM$ as
\[ v = \dot{a}(t) \, u(t) + a(t) \, \sum_{\ell=1}^{N} \varphi_1(t) \wedge \cdots \wedge \varphi_{\ell-1}(t) \wedge \theta_\ell \wedge \varphi_{\ell+1}(t) \wedge \cdots \wedge \varphi_N(t) \]
where $\theta_\ell \in L^2(\Lambda)$ is any function, possibly orthogonal to $\set{\varphi_1(t), \ldots, \varphi_N(t)}$ in view of Proposition~\ref{prop:GaugeFreedom}. The above yields then an orthogonal decomposition of the vector $v$ in~$\cH_N$. Therefore, Equations~\eqref{eqn:dota} and~\eqref{eqn:HF} will be derived by plugging in~\eqref{condDirac} each of the orthogonal summands
\begin{equation} \label{eqn:vuvi}
v_u = u(t) \quad \text{and} \quad v_\ell = \varphi_1(t) \wedge \cdots \wedge \varphi_{\ell-1}(t) \wedge \theta_\ell \wedge \varphi_{\ell+1}(t) \wedge \cdots \wedge \varphi_N(t)\,.
\end{equation}

If the orbitals $\set{\varphi_1(t), \ldots, \varphi_N(t)}$ are chosen so that the condition in~\eqref{eqn:GaugeFreedom} holds, then by taking the scalar product of~\eqref{equdot} with $u(t) = a(t) \, \varphi_1(t) \wedge \cdots \wedge \varphi_N(t)$ we get
\[ \scal{u(t)}{\dot{u}(t)} = \dot{a}(t) \, \overline{a(t)} \, \scal{\varphi_1(t) \wedge \cdots \wedge \varphi_N(t)}{\varphi_1(t) \wedge \cdots \wedge \varphi_N(t)} = \dot{a}(t) \, \overline{a(t)} \]
and by using~\eqref{eqn:|a|}, i.e.\ $\overline{a(t)} = a(t)^{-1}$, we arrive at 
\[ \dot{a}(t) = \scal{u(t)}{\dot{u}(t)} \, a(t)\,. \]
Using the fact that $u(t)$ is the solution to~\eqref{condDirac}, we have now
\[ \scal{u(t)}{\dot{u}(t)} = \frac{1}{\iu \, \hbar} \, \scal{u(t)}{H_N \, u(t)} \equiv \frac{1}{\iu \, \hbar} \, \scal{u(0)}{H_N \, u(0)} = \frac{1}{\iu \, \hbar} \, \cE_0^{(N)}\,, \]
in view of Proposition~\ref{prop:Preservation}. Combining the two equalities above yields~\eqref{eqn:dota}, as wanted. 

To derive now the Hartree--Fock equations, let us consider~\eqref{condDirac} with $v = v_\ell$ as in~\eqref{eqn:vuvi} and $\dot{u}(t)$ as in~\eqref{equdot}, assuming the gauge condition~\eqref{eqn:GaugeFreedom}. As was already mentioned, only the part of the function $\theta_\ell$ which is orthogonal to $\varphi_1(t), \ldots, \varphi_N(t)$ contributes to the scalar product on the right-hand side of~\eqref{condDirac}. With this in mind, the Dirac--Frenkel condition reads
\begin{equation} \label{eqn:theta|HF}
\begin{aligned}
0 & = \scal{\theta_\ell}{\iu \hbar \, a(t) \, \dot{\varphi}_\ell(t)} - \scal{\varphi_1(t) \wedge \cdots \wedge \varphi_{\ell-1}(t) \wedge \theta_\ell \wedge \varphi_{\ell+1}(t) \wedge \cdots \wedge \varphi_N(t)}{H_N\,u(t)} \\
& = a(t) \, \big[ \scal{\theta_\ell}{\iu \hbar \, \dot{\varphi}_\ell(t)} \\
& \quad - \scal{\varphi_1(t) \wedge \cdots \wedge \varphi_{\ell-1}(t) \wedge \theta_\ell \wedge \varphi_{\ell+1}(t) \wedge \cdots \wedge \varphi_N(t)}{H_N\, \varphi_1(t) \wedge \cdots \wedge \varphi_N(t)} \big] 
\end{aligned}
\end{equation}
for all $\theta_\ell \in L^2(\Lambda)$, which at this stage we can assume to be normalized. To compute the term on the last line of the above, we use the form~\eqref{eqn:accaenne} of $H_N$ and the orthonormality conditions. From~\eqref{eqn:ScalarSlater} we have
\begin{equation} \label{eqn:HFni}
\begin{aligned}
& \scal{\varphi_1(t)  \wedge \cdots \wedge \varphi_{\ell-1}(t) \wedge \theta_\ell \wedge \varphi_{\ell+1}(t) \wedge \cdots \wedge \varphi_N(t)}{{H}\sub{ni}\, \varphi_1(t) \wedge \cdots \wedge \varphi_N(t)} \\
& \quad = \sum_{1 \le i \le N} \scal{\varphi_1(t) \wedge \cdots \wedge \theta_\ell \wedge \cdots \wedge \varphi_N(t)}{\varphi_1(t) \wedge \cdots \wedge \big( H_1 \, \varphi_i(t) \big) \wedge \cdots \wedge \varphi_N(t)} \\
& \quad = \sum_{1 \le i \le N} \delta_{i,\ell} \scal{\theta_\ell}{H_1 \, \varphi_i(t)} = \scal{\theta_\ell}{H_1\, \varphi_\ell(t)}\,.
\end{aligned}
\end{equation}
We claim that the interaction term in $H_N$ yields
\begin{equation} \label{eqn:interac}
\begin{aligned}
& \scal{\varphi_1(t)  \wedge \cdots \wedge \varphi_{\ell-1}(t) \wedge \theta_\ell \wedge \varphi_{\ell+1}(t) \wedge \cdots \wedge \varphi_N(t)}{H\sub{int}\, \varphi_1(t) \wedge \cdots \wedge \varphi_N(t)} \\
& \quad = \sum_{\ell' \ne \ell} \scal{\theta_\ell \otimes \varphi_{\ell'}(t)}{V_{12} \, \varphi_\ell(t) \otimes \varphi_{\ell'}(t)} - \scal{\theta_\ell \otimes \varphi_{\ell'}(t)}{V_{12} \, \varphi_{\ell'}(t) \otimes \varphi_{\ell}(t)}
\end{aligned}
\end{equation}
where $V_{12}$ denotes the multiplication operator by $V(\vec{x};\vec{y})$ on $L^2(\Lambda)\otimes L^2(\Lambda)$. 
We will prove the previous claim momentarily; let us first show how this identity determines the definition of the Hartree--Fock and exchange potentials~\eqref{eqn:KHF} and~\eqref{eqn:XHF}. Indeed, the right-hand side of~\eqref{eqn:interac} can be spelled out to be
\begin{equation} \label{eqn:HFv}
\begin{aligned}
\sum_{\ell' \ne \ell} & \int_{\Lambda} \di \vec{x} \, \int_{\Lambda} \di \vec{y} \; \overline{\theta_\ell(\vec{x})} \, \overline{\varphi_{\ell'}(t,\vec{y})} \, V(\vec{x};\vec{y}) \, \left[ \varphi_\ell(t,\vec{x}) \, \varphi_{\ell'}(t,\vec{y}) - \varphi_{\ell'}(t,\vec{x}) \, \varphi_{\ell}(t,\vec{y}) \right] \\
& = \int_{\Lambda} \di \vec{x} \, \overline{\theta_\ell(\vec{x})} \, \left[ \sum_{\ell' \ne \ell} \int_{\Lambda} \di \vec{y} \; V(\vec{x};\vec{y}) \, \left|\varphi_{\ell'}(t,\vec{y})\right|^2 \right] \, \varphi_\ell(t,\vec{x}) \\
& \quad - \sum_{\ell' \ne \ell} \int_{\Lambda} \di \vec{x} \, \overline{\theta_\ell(\vec{x})} \, \left[ \sum_{\ell' \ne \ell} \int_{\Lambda} \di \vec{y} \; V(\vec{x};\vec{y}) \, \overline{\varphi_{\ell'}(t,\vec{y})} \, \varphi_{\ell}(t,\vec{y}) \right] \, \varphi_{\ell'}(t,\vec{x}) \\
& = \scal{\theta_\ell}{K_\ell(t) \, \varphi_\ell(t) - \sum_{\ell'\ne \ell} X_{\ell,\ell'}(t) \, \varphi_{\ell'}(t)}\,.
\end{aligned}
\end{equation}
Plugging~\eqref{eqn:HFni} and~\eqref{eqn:HFv} into~\eqref{eqn:theta|HF} yields
\[ \scal{\theta_\ell}{\iu \hbar \, \dot{\varphi}_\ell(t) - H_1 \, \varphi_\ell(t) - K_\ell(t) \, \varphi_\ell(t) + \sum_{\ell'\ne \ell} X_{\ell,\ell'}(t) \, \varphi_{\ell'}(t) } = 0 \quad \text{for all } \theta_\ell \in L^2(\Lambda)\,, \]
and therefore the Hartree--Fock equations~\eqref{eqn:HF} hold, as claimed.

It remains to show~\eqref{eqn:interac}. Let us first rewrite the definition~\eqref{eqn:wedgedef} of the wedge product as
\begin{equation} \label{eqn:detensor}
\psi_1 \wedge \cdots \wedge \psi_N = \frac{1}{\sqrt{N!}}\, \sum_{\sigma \in \mathfrak{S}_N} (-1)^\sigma \, \psi_{\sigma(1)} \otimes \cdots \otimes \psi_{\sigma(N)}\,,
\end{equation}
where $\mathfrak{S}_N$ is the permutation group on $N$ elements, and $(-1)^\sigma$ is the parity of a permutation $\sigma \in \mathfrak{S}_N$. It is also convenient to write the function $V \in L^\infty(\Lambda \times \Lambda) \subset L^2(\Lambda \times \Lambda) \simeq L^2(\Lambda) \otimes L^2(\Lambda)$ as
\[ V = \sum_{\alpha, \beta \in \N \times \Z/M\Z} v_{\alpha \beta} \, \varphi_\alpha \otimes \varphi_\beta\,, \]
where $\set{\varphi_\alpha \otimes \varphi_\beta \equiv \varphi_{n,m} \otimes \varphi_{n',m'}}_{\alpha = (n,m), \beta = (n',m') \in \N \times \Z / M \Z}$ is the orthonormal basis of the tensor product $L^2(\Lambda) \otimes L^2(\Lambda)$ constructed from the eigenfunctions of the Landau Hamiltonian $H_1$ (compare Theorem~\ref{thm:Landau1}). Notice that the symmetry $V(\vec{x};\vec{y}) = V(\vec{y};\vec{x})$ of $V$ yields
\begin{equation} \label{eqn:valphabeta}
V = \sum_{\alpha, \beta \in \N \times \Z/M\Z} v_{\alpha \beta} \, \varphi_\alpha \otimes \varphi_\beta = \sum_{\alpha, \beta \in \N \times \Z/M\Z} v_{\alpha \beta} \, \varphi_\beta \otimes \varphi_\alpha \,.
\end{equation}
Then, if $V_{ij}$ is the operator of multiplication by $V$ on the space of the $i$-th and $j$-th particles, $i<j \in \set{1,\ldots,N}$, as in~\eqref{eqn:hatV}, we have
\begin{equation} \label{eqn:VijSlater}
\begin{aligned}
V_{ij}\, & \varphi_1 \wedge \cdots \wedge \varphi_N = \frac{1}{\sqrt{N!}}\, \sum_{\sigma \in \mathfrak{S}_N} (-1)^\sigma \, V_{ij} \, \varphi_{\sigma(1)} \otimes \cdots \otimes \varphi_{\sigma(N)} \\
& = \frac{1}{\sqrt{N!}}\, \sum_{\sigma \in \mathfrak{S}_N} (-1)^\sigma \\
& \qquad \sum_{\alpha, \beta \in \N \times \Z/M\Z} v_{\alpha \beta} \, \varphi_{\sigma(1)} \otimes \cdots \otimes \left( \varphi_\alpha \, \varphi_{\sigma(i)} \right) \otimes \cdots \otimes \left( \varphi_\beta \, \varphi_{\sigma(j)} \right) \otimes \cdots \otimes \varphi_{\sigma(N)}\,,
\end{aligned}
\end{equation}
where we dropped the dependence on $t$ of $\varphi_1(t), \ldots, \varphi_N(t)$ for notational convenience. Let us take the scalar product of the above with $\varphi_1 \wedge \cdots \wedge \theta_\ell \wedge \varphi_N$. This yields, using again~\eqref{eqn:detensor}, 
\begin{equation} \label{Vij product}
\begin{aligned}
& \scal{\varphi_1 \wedge \cdots \wedge \theta_\ell \wedge \varphi_N}{V_{ij}\,  \varphi_1 \wedge \cdots \wedge \varphi_N} \\
& \quad = \frac{1}{N!}\, \sum_{\sigma, \sigma' \in \mathfrak{S}_N} (-1)^{\sigma'} \, (-1)^\sigma \, \sum_{\alpha, \beta \in \N \times \Z/M\Z} v_{\alpha \beta} \, \\
& \qquad \scal{\varphi'_{\sigma'(1)} \otimes \cdots \otimes \varphi'_{\sigma'(N)}}{\varphi_{\sigma(1)} \otimes \cdots \otimes \left( \varphi_\alpha \, \varphi_{\sigma(i)} \right) \otimes \cdots \otimes \left( \varphi_\beta \, \varphi_{\sigma(j)} \right) \otimes \cdots \otimes \varphi_{\sigma(N)}} \\
& \quad = \frac{1}{N!}\, \sum_{\sigma, \sigma' \in \mathfrak{S}_N} (-1)^{\sigma'} \, (-1)^\sigma \, \sum_{\alpha, \beta \in \N \times \Z/M\Z} v_{\alpha \beta} \, \\
& \qquad \scal{\varphi'_{\sigma'(1)}}{\varphi_{\sigma(1)}} \, \cdots \, \scal{\varphi'_{\sigma'(i)}}{\varphi_\alpha \, \varphi_{\sigma(i)}} \, \cdots \, \scal{\varphi'_{\sigma'(j)}}{\varphi_\beta \, \varphi_{\sigma(j)}} \, \cdots \, \scal{\varphi'_{\sigma'(N)}}{\varphi_{\sigma(N)}}\,,
\end{aligned}
\end{equation} 
where for $\ell' \in \set{1, \ldots, N}$
\[ \varphi'_{\ell'} := \begin{cases} \varphi_{\ell'} & \text{if } \ell' \ne \ell\,, \\ \theta_\ell & \text{if } \ell' = \ell\,. \end{cases} \]

Let us now analyze the right-hand side of the above equality. If the permutation $\sigma' \in \mathfrak{S}_N$ is such that $\sigma'(i) \ne \ell \ne \sigma'(j)$, then $\ell = \sigma'(k)$ for some $i \ne k \ne j$, and thefore the product of scalar products in the right-hand side of the above equality contains the factor $\scal{\theta_\ell}{\varphi_{\sigma(k)}} = 0$. Therefore, the only permutations which contribute to the sum are those for which either $\sigma'(i) = \ell$ or $\sigma'(j) = \ell$. Let us fix $\alpha, \beta \in \N \times \Z / M \Z$ and look at one summand of the corresponding sum, for which we have that the only non-zero contributions are
\begin{align*}
& \sum_{\substack{\sigma' \in \mathfrak{S}_N \\ \sigma'(i) = \ell}} (-1)^{\sigma'} \sum_{\sigma \in \mathfrak{S}_N} (-1)^\sigma \, \delta_{\sigma'(1), \, \sigma(1)} \, \cdots \, \scal{\theta_\ell}{\varphi_\alpha \, \varphi_{\sigma(i)}} \, \cdots \, \scal{\varphi_{\sigma'(j)}}{\varphi_\beta \, \varphi_{\sigma(j)}} \, \cdots \, \delta_{\sigma'(N),\,\sigma(N)} \\
& + \sum_{\substack{\sigma' \in \mathfrak{S}_N \\ \sigma'(j) = \ell}} (-1)^{\sigma'} \sum_{\sigma \in \mathfrak{S}_N} (-1)^\sigma \, \delta_{\sigma'(1), \, \sigma(1)} \, \cdots \, \scal{\varphi_{\sigma'(i)}}{\varphi_\alpha \, \varphi_{\sigma(i)}} \, \cdots \, \scal{\theta_\ell}{\varphi_\beta \, \varphi_{\sigma(j)}} \, \cdots \, \delta_{\sigma'(N),\,\sigma(N)}\,.
\end{align*}

Let us focus on the first sum above. If $\sigma'(i) = \ell$, and we call $\sigma'(j) =: \ell'$, then the only two permutations $\sigma \in \mathfrak{S}_N$ which give a non-zero contribution to the sum are $\sigma = \sigma'$ or $\sigma = \sigma' \circ (ij)$; namely, $\sigma$ can be the permutation coinciding with $\sigma'$ --- for which $(-1)^{\sigma} = (-1)^{\sigma'}$ --- or the one which coincides with $\sigma'$ apart from at $i$ and $j$, where instead we have $\sigma(i) = \ell'$ and $\sigma(j) = \ell$ --- for which $(-1)^{\sigma} = (-1)^{\sigma'} \, (-1)^{(ij)} = -(-1)^{\sigma'}$. A similar argument can be made for the sum over permutations $\sigma'$ such that $\sigma'(j) = \ell$. All in all, the above sums equal
\begin{align*}
& \sum_{\substack{\sigma' \in \mathfrak{S}_N \\ \sigma'(i) = \ell}} \big[ \scal{\theta_\ell}{\varphi_\alpha \, \varphi_{\ell}} \, \cdot \, \scal{\varphi_{\sigma'(j)}}{\varphi_\beta \, \varphi_{\sigma'(j)}} - \scal{\theta_\ell}{\varphi_\alpha \, \varphi_{\sigma'(j)}} \, \cdot \, \scal{\varphi_{\sigma'(j)}}{\varphi_\beta \, \varphi_{\ell}} \big] \\
& + \sum_{\substack{\sigma' \in \mathfrak{S}_N \\ \sigma'(j) = \ell}} \big[ \scal{\varphi_{\sigma'(i)}}{\varphi_\alpha \, \varphi_{\sigma'(i)}} \, \cdot \, \scal{\theta_\ell}{\varphi_\beta \, \varphi_{\ell}} - \scal{\varphi_{\sigma'(i)}}{\varphi_\alpha \, \varphi_{\ell}} \, \cdot \, \scal{\theta_\ell}{\varphi_\beta \, \varphi_{\sigma'(i)}} \big]\,.
\end{align*} 

We now write
\begin{gather*}
\set{\sigma' \in \mathfrak{S}_N : \sigma'(i) = \ell} = \bigsqcup_{\ell' \ne \ell} \set{\sigma' \in \mathfrak{S}_N : \sigma'(i) = \ell \text{ and } \sigma'(j) = \ell'} \,, \\
\set{\sigma' \in \mathfrak{S}_N : \sigma'(j) = \ell} = \bigsqcup_{\ell' \ne \ell} \set{\sigma' \in \mathfrak{S}_N : \sigma'(j) = \ell \text{ and } \sigma'(i) = \ell'} \,,
\end{gather*}
and notice that
\begin{align*}
\sharp \set{\sigma' \in \mathfrak{S}_N : \sigma'(i) = \ell \text{ and } \sigma'(j) = \ell'} & = (N-2)! \\
& = \sharp \set{\sigma' \in \mathfrak{S}_N : \sigma'(j) = \ell \text{ and } \sigma'(i) = \ell'} \,.
\end{align*}
With these identifications, it is easy to realize that the above sums compute
\begin{align*}
(N-2)! \, \sum_{\ell' \ne \ell} & \scal{\theta_\ell}{\varphi_\alpha \, \varphi_{\ell}} \, \cdot \, \scal{\varphi_{\ell'}}{\varphi_\beta \, \varphi_{\ell'}} - \scal{\theta_\ell}{\varphi_\alpha \, \varphi_{\ell'}} \, \cdot \, \scal{\varphi_{\ell'}}{\varphi_\beta \, \varphi_{\ell}} \\
& + \scal{\varphi_{\ell'}}{\varphi_\alpha \, \varphi_{\ell'}} \, \cdot \, \scal{\theta_\ell}{\varphi_\beta \, \varphi_{\ell}} - \scal{\varphi_{\ell'}}{\varphi_\alpha \, \varphi_{\ell}} \, \cdot \, \scal{\theta_\ell}{\varphi_\beta \, \varphi_{\ell'}} \\
= (N-2)! \, \sum_{\ell' \ne \ell} & \scal{\theta_\ell \otimes \varphi_{\ell'}}{\varphi_\alpha \, \varphi_{\ell} \, \otimes \, \varphi_\beta \, \varphi_{\ell'}} - \scal{\theta_\ell \otimes \varphi_{\ell'}}{\varphi_\alpha \, \varphi_{\ell'} \, \otimes \, \varphi_\beta \, \varphi_{\ell}} \\
& + \scal{\theta_\ell \otimes \varphi_{\ell'}}{\varphi_\beta \, \varphi_{\ell} \, \otimes \, \varphi_\alpha \, \varphi_{\ell'}} - \scal{\theta_\ell \otimes \varphi_{\ell'}}{\varphi_\beta \, \varphi_{\ell'} \, \otimes \, \varphi_\alpha \, \varphi_{\ell}} \,.
\end{align*}

Let us plug the above information back in~\eqref{Vij product}: so we multiply by $v_{\alpha \beta}$, sum over $\alpha, \beta \in \N \times \Z / M \Z$ and divide by $N!$. Upon using the symmetry condition~\eqref{eqn:valphabeta}, this whole expression simplifies to
\begin{align*}
& \scal{\varphi_1 \wedge \cdots \wedge \theta_\ell \wedge \varphi_N}{V_{ij}\,  \varphi_1 \wedge \cdots \wedge \varphi_N} \\
& \quad = \frac{1}{N!} \cdot (N-2)! \cdot 2 \, \sum_{\ell' \ne \ell} \scal{\theta_\ell \otimes \varphi_{\ell'}}{V \, \varphi_{\ell} \otimes \varphi_{\ell'}} - \scal{\theta_\ell \otimes \varphi_{\ell'}}{V \, \varphi_{\ell'} \otimes \varphi_{\ell}} \\
& \quad = \frac{2}{N\,(N-1)} \, \sum_{\ell' \ne \ell} \scal{\theta_\ell \otimes \varphi_{\ell'}}{V_{12} \, \varphi_{\ell} \otimes \varphi_{\ell'}} - \scal{\theta_\ell \otimes \varphi_{\ell'}}{V_{12} \, \varphi_{\ell'} \otimes \varphi_{\ell}} \,. 
\end{align*}
The right-hand side of the above is independent of $i<j \in \set{1,\ldots,N}$, and therefore summing over the $\binom{N}{2} = N \, (N-1) / 2$ choices of such indices we conclude that
\begin{multline*}
\sum_{1 \le i < j \le N} \scal{\varphi_1 \wedge \cdots \wedge \theta_\ell \wedge \varphi_N}{V_{ij}\,  \varphi_1 \wedge \cdots \wedge \varphi_N} \\
= \sum_{\ell' \ne \ell} \scal{\theta_\ell \otimes \varphi_{\ell'}}{V_{12} \, \varphi_{\ell} \otimes \varphi_{\ell'}} - \scal{\theta_\ell \otimes \varphi_{\ell'}}{V_{12} \, \varphi_{\ell'} \otimes \varphi_{\ell}}
\end{multline*}
which coincides with~\eqref{eqn:interac}.

Now that the Hartree--Fock equations~\eqref{eqn:HF} have been established, the regularity properties of the solutions stem from the theory of nonlinear partial differential equations. The interested reader is referred to~\cite{Lubich08}; see also~\cite{CazenaveEsteban88} and~\cite[Section 9.1]{Cazenave03} for related works.
\end{proof}

\section{Effectiveness of the Hartree--Fock dynamics} \label{sec:Estimate}

Having established the Hartree--Fock equations, we want to provide a quantitative estimate on how good of an approximation the solution $u(t)$ of the Dirac--Frenkel principle~\eqref{condDirac} is compared to the solution $\Psi(t) = \eu^{-\iu \, H_N \, t / \hbar}\, \Psi(0)$ to the $N$-body Schr\"odinger equation, provided both share the same initial condition $u(0) = \Psi(0)$. For simplicity and without loss of generality we assume
\[ u(0) \equiv \Psi(0) = \Phi\sub{Slater}(0) = \varphi_1(0) \wedge \cdots \wedge \varphi_N(0) \in \bigwedge_{i=1}^{N} H^2_{\vec{A}}(\Lambda)\,, \]
i.e.\ $a(0) = 1$ in the representation~\eqref{varietàHF} of $u(0) \in \cM$: this phase can be reabsorbed in the definition of one of the orbitals $\varphi_\ell(0)$.

\begin{theorem} \label{thm:Estimate}
Let $u(t) = a(t) \, \varphi_1(t) \wedge \cdots \wedge \varphi_N(t) \in \cM$, $t \in [0,\overline{t}]$, be as in Theorem~\ref{thm:HF}, with $a(0) = 1$. Let also $\Psi(t) = \eu^{-\iu \, H_N \, t / \hbar}\, \Phi\sub{Slater}(0)$ be the solution of the $N$-body Schr\"odinger equation
\[ \iu \, \hbar \, \dot{\Psi}(t) = H_N \, \Psi(t)\,. \]
Then for all $t \in [0,\overline{t}]$
\begin{equation} \label{eqn:Estimate}
\norm{\Psi(t) -u(t)}_{\cH_N} \le \frac{1}{\hbar}\, \sqrt{N(N-1)} \, \norm{V}_{L^\infty(\Lambda \times \Lambda)} \, t\,.
\end{equation}
\end{theorem}

\begin{proof}
Let us compute then, for $s \in [0,t] \subset [0,\overline{t}]$,
\begin{align*}
\norm{\Psi(s) - u(s)}&  \cdot \frac{\di}{\di s} \norm{\Psi(s) - u(s)} = \frac{1}{2} \, \frac{\di}{\di s} \norm{\Psi(s) - u(s)}^2 \\
& = \re \scal{\Psi(s) - u(s)}{\dot{\Psi}(s) - \dot{u}(s)} \\
& = \re \scal{\Psi(s) - u(s)}{\frac{1}{\iu \hbar} \, H_N \, \left[ \Psi(s) - u(s) \right]} \\
& \quad - \re \scal{\Psi(s) - u(s)}{\dot{u}(s) - \frac{1}{\iu \hbar} \, H_N \,u(s)}\,.
\end{align*}
The first term on the right-hand side of the above equality vanishes, because $H_N$ is self-adjoint. Thefore, by applying the Cauchy--Schwarz inequality to the scalar product in the second term and dividing both sides by $\norm{\Psi(s) - u(s)}$, we conclude that
\[ \left| \frac{\di}{\di s} \norm{\Psi(s) - u(s)} \right| \le \norm{\dot{u}(s) - \frac{1}{\iu \hbar} \, H_N \,u(s)}\,. \]
With the above, by the fundamental theorem of calculus we can estimate for $t \in [0,\overline{t}]$
\begin{equation} \label{eqn:APosteriori}
\norm{\Psi(t) - u(t)} \le \int_{0}^{t} \di s \, \norm{\dot{u}(s) - \frac{1}{\iu \hbar} \, H_N \,u(s)}
\end{equation}
(compare e.g.~\cite[Chapter~II, Theorem~1.5]{Lubich08}).

We now bound the integrand on the right-hand side of the above. To this end, let us notice that
\begin{align*}
\dot{u}(s) & = \frac{\di}{\di s} \left[ a(s) \, \varphi_1(s) \wedge \cdots \wedge \varphi_N(s) \right] \\
& = \dot{a}(s) \, \varphi_1(s) \wedge \cdots \wedge \varphi_N(s) + a(s) \sum_{\ell=1}^{N} \varphi_1(s) \wedge \cdots \wedge \dot{\varphi}_\ell(s) \wedge \cdots \wedge \varphi_N(s) \\
&  = \frac{\cE_0^{(N)}}{\iu \hbar} \, a(s) \, \varphi_1(s) \wedge \cdots \wedge \varphi_N(s) + a(s) \, \sum_{\ell=1}^{N} \varphi_1(s) \wedge \cdots \\
& \qquad \cdots \wedge \left\{\frac{1}{\iu \hbar} \, \left[ H_1 \, \varphi_\ell(s) + K_\ell(s) \varphi_\ell(s) - \sum_{\ell'\ne\ell} X_{\ell,\ell'}(s) \, \varphi_{\ell'}(s) \right] \right\} \wedge \cdots \wedge \varphi_N(s)
\end{align*}
in view of~\eqref{eqn:dota} and the Hartree--Fock equations~\eqref{eqn:HF}. Similarly
\begin{align*}
H_N \, u(s) & = a(s) \, \left({H}\sub{ni} + H\sub{int} \right)\, \left[ \varphi_1(s) \wedge \cdots \wedge \varphi_N(s) \right] \\
& = a(s) \, \left[ \sum_{\ell=1}^{N} \varphi_1(s) \wedge \cdots \wedge H_1 \, \varphi_\ell(s) \wedge \cdots \wedge \varphi_N(s) \right] + a(s) \, H\sub{int} \, \left[ \varphi_1(s) \wedge \cdots \wedge \varphi_N(s) \right]
\end{align*}
and therefore
\begin{equation} \label{eqn:udot-Hu}
\begin{aligned} 
\dot{u}(s) & - \frac{1}{\iu \hbar} \, H_N \,u(s) = \frac{\cE_0^{(N)}}{\iu \hbar} \, u(s) \\
& + \frac{a(s)}{\iu \hbar} \Bigg\{ \left[ \sum_{\ell=1}^{N} \varphi_1(s) \wedge \cdots \wedge \left( K_\ell(s) \varphi_\ell(s) - \sum_{\ell'\ne\ell} X_{\ell,\ell'}(s) \, \varphi_{\ell'}(s) \right) \wedge \cdots \wedge \varphi_N(s) \right] \\
& \quad - H\sub{int} \, \left[ \varphi_1(s) \wedge \cdots \wedge \varphi_N(s) \right] \Bigg\}\,.
\end{aligned}
\end{equation}

In order to compute the norm of the left-hand side of the above, which appears in~\eqref{eqn:APosteriori}, we use the fact that
\begin{equation} \label{eqn:dualnorm}
\norm{\dot{u}(s) - \frac{1}{\iu \hbar} \, H_N \,u(s)} = \sup_{\substack{\Phi \in \cH_N \\ \norm{\Phi} = 1}} \left| \scal{\Phi}{\dot{u}(s) - \frac{1}{\iu \hbar} \, H_N \,u(s)} \right| \,.
\end{equation}
To estimate the absolute value of the scalar product appearing on the right-hand side, let us decompose $\cH_N = \bigwedge_{i=1}^{N} L^2(\Lambda)$ into an orthogonal sum of subspaces each accounting for a certain number of orbitals out of the Hartree--Fock Slater determinant, i.e.,
\begin{equation} \label{eqn:decompo}
\cH_N = \cH_N^{(0)} \oplus \cH_N^{(1)} \oplus \cH_N^{(2)} \oplus \cdots \oplus \cH_N^{(N)}\,,
\end{equation}
where the summands are defined as follows:
\begin{itemize}
 \item $\cH_N^{(0)}$ is spanned by the Slater determinant $\varphi_1(s) \wedge \cdots \wedge \varphi_N(s)$;
 \item $\cH_N^{(1)}$ is spanned by Slater determinants of the form
 \[ \varphi_1(s) \wedge \cdots \wedge \varphi_{m-1}(s) \wedge \theta_m \wedge \varphi_{m+1}(s) \wedge \cdots \wedge \varphi_N(s)\,, \quad m \in \set{1, \ldots, N}\,, \]
 where $\theta_m \in L^2(\Lambda)$ is a (normalized) function orthogonal to $\set{\varphi_1(s), \ldots, \varphi_N(s)}$;
  \item $\cH_N^{(2)}$ is spanned by Slater determinants of the form
 \[ \varphi_1(s) \wedge \cdots \wedge \theta_{m_1} \wedge \cdots \wedge \theta_{m_2} \wedge \cdots \wedge \varphi_N(s)\,, \quad m_1 < m_2 \in \set{1, \ldots, N}\,, \]
 where $\theta_{m_1}, \theta_{m_2} \in L^2(\Lambda)$ are (normalized) functions orthogonal to $\set{\varphi_1(s), \ldots, \varphi_N(s)}$ and to each other;
 \item for $i \in \set{1, \ldots, N}$, the subspace $\cH_N^{(i)}$ is spanned by Slater determinants in which~$i$ of the $N$ orbitals $\varphi_1(s), \ldots, \varphi_N(s)$ have been swapped for orthogonal functions $\theta_{m_1}, \ldots, \theta_{m_i}$ (for all possible choices of indices $m_1 < \cdots < m_i$ in $\set{1, \ldots, N}$) which are orthogonal to all the orbitals $\varphi_1(s), \ldots, \varphi_N(s)$.
\end{itemize}
Notice that the subspaces $\cH_N^{(i)}$ are indeed mutually orthogonal, in view of~\eqref{eqn:ScalarSlater}, and that they generate the whole $\cH_N$. Their definition depends in principle on $s \in [0,t]$, but we will not keep track of it in the notation.

Each vector $\Phi \in \cH_N$ can be decomposed orthogonally as the sum of its components $\Phi^{(i)} \in \cH_N^{(i)}$, according to~\eqref{eqn:decompo}, and therefore
\[ \scal{\Phi}{\dot{u}(s) - \frac{1}{\iu \hbar} \, H_N \,u(s)} = \sum_{i=0}^{N} \scal{\Phi^{(i)}}{\dot{u}(s) - \frac{1}{\iu \hbar} \, H_N \,u(s)} \,. \]
To compute the scalar products on the right for a generic $\Phi^{(i)}$, it clearly suffices to compute them on the generating Slater determinants $\Phi^{(i)}\sub{gen}$ of $\cH_N^{(i)}$.
\begin{itemize}
 \item On $\cH_N^{(0)}$, we have
 \begin{equation} \label{eqn:HF0thorder}
 \scal{\varphi_1(s) \wedge \cdots \wedge \varphi_N(s)}{\dot{u}(s) - \frac{1}{\iu \hbar} \, H_N \,u(s)} = a(s) \, \scal{u(s)}{\dot{u}(s) - \frac{1}{\iu \hbar} \, H_N \,u(s)} = 0 
 \end{equation}
 in view of~\eqref{condDirac} and the fact that $u(s) \in T_{u(s)} \cM$, as $\cM$ contains rays.
 \item Let us now pick $m \in \set{1, \ldots, N}$ and $\Phi^{(1)}\sub{gen} = \varphi_1(s) \wedge \cdots \wedge \theta_m \wedge \cdots \wedge \varphi_N(s) \in \cH_N^{(1)}$, with $\theta_m \in L^2(\Lambda)$ orthogonal to all the $\varphi_i(s)$'s. In particular, $\Phi^{(1)}\sub{gen}$ is orthogonal to $u(s) = a(s) \, \varphi_1(s) \wedge \cdots \wedge \varphi_N(s) \in \cH_N^{(0)}$, and therefore from~\eqref{eqn:udot-Hu}
 \begin{equation} \label{eqn:Phi(1)}
 \begin{aligned}
\scal{\Phi^{(1)}\sub{gen}}{\dot{u}(s) - \frac{1}{\iu \hbar} \, H_N \,u(s)} = \frac{a(s)}{\iu \hbar} \Bigg\{ & \left[ \sum_{\ell=1}^{N} \scal{\Phi^{(1)}\sub{gen}}{\varphi_1(s) \wedge \cdots \wedge \eta_\ell(s) \wedge \cdots \wedge \varphi_N(s)} \right] \\
& \quad - \scal{\Phi^{(1)}\sub{gen}}{H\sub{int} \, \left[ \varphi_1(s) \wedge \cdots \wedge \varphi_N(s) \right]} \Bigg\}\,,
 \end{aligned}
 \end{equation}
 where also for future convenience we have denoted by $\eta_\ell(s)$ the nonlinear part of the $\ell$-th Hartree--Fock equation:
 \[ \eta_\ell(s) \equiv \eta_\ell(s; \varphi_1(s), \ldots, \varphi_N(s)) := K_\ell(s) \varphi_\ell(s) - \sum_{\ell'\ne\ell} X_{\ell,\ell'}(s) \, \varphi_{\ell'}(s)\,. \]
 
 From~\eqref{eqn:ScalarSlater} and the orthogonality conditions on $\varphi_1(s), \ldots, \varphi_N(s)$ and $\theta_m$, we see at once that
 \begin{align*}
 & \scal{\Phi^{(1)}\sub{gen}}{\varphi_1(s) \wedge \cdots \wedge \eta_\ell(s) \wedge \cdots \wedge \varphi_N(s)} \\
 & = \scal{\varphi_1(s) \wedge \cdots \wedge \theta_m \wedge \cdots \wedge \varphi_N(s)}{\varphi_1(s) \wedge \cdots \wedge \eta_\ell(s) \wedge \cdots \wedge \varphi_N(s)} \\
 & = \delta_{\ell, m} \, \scal{\theta_m}{\eta_\ell(s)}\,.
 \end{align*}
 Notice also that~\eqref{eqn:interac} and~\eqref{eqn:HFv} together yield
 \begin{align*}
 & \scal{\Phi^{(1)}\sub{gen}}{H\sub{int} \, \left[ \varphi_1(s) \wedge \cdots \wedge \varphi_N(s) \right]} \\
 & = \scal{\varphi_1(s) \wedge \cdots \wedge \theta_m \wedge \cdots \wedge \varphi_N(s)}{H\sub{int} \, \left[ \varphi_1(s) \wedge \cdots \wedge \varphi_N(s) \right]} \\
& = \scal{\theta_m}{\eta_m(s)}\,.
 \end{align*}
 Plugging the above two equalities in~\eqref{eqn:Phi(1)} we conclude that
 \begin{equation} \label{eqn:HF1storder}
 \scal{\Phi^{(1)}\sub{gen}}{\dot{u}(s) - \frac{1}{\iu \hbar} \, H_N \,u(s)} = 0
 \end{equation}
 for the generators $\Phi^{(1)}\sub{gen}$ of $\cH_N^{(1)}$; therefore, all $\Phi^{(1)}$ in $\cH_N^{(1)}$ satisfy the same orthogonality condition.
 
 \item We now choose a generator $\Phi^{(2)}\sub{gen} = \varphi_1(s) \wedge \cdots \wedge \theta_{m_1} \wedge \cdots \wedge \theta_{m_2} \wedge \cdots \wedge \varphi_N(s)$ of $\cH_N^{(2)}$. Since both $\theta_{m_1}$ and $\theta_{m_2}$ are orthogonal to $\varphi_1(s), \ldots, \varphi_N(s)$, it follows from~\eqref{eqn:ScalarSlater} that
 \begin{gather*}
 \scal{\Phi^{(2)}\sub{gen}}{u(s)} = a(s) \, \scal{\Phi^{(2)}\sub{gen}}{\varphi_1(s) \wedge \cdots \wedge \varphi_N(s)} = 0\,, \\
 \scal{\Phi^{(2)}\sub{gen}}{\varphi_1(s) \wedge \cdots \wedge \eta_\ell(s) \wedge \cdots \wedge \varphi_N(s)} = 0\,,
 \end{gather*}
 and therefore, using~\eqref{eqn:udot-Hu}, that
 \begin{equation} \label{eqn:scalPhi(2)}
 \begin{aligned}
 & \scal{\Phi^{(2)}\sub{gen}}{\dot{u}(s) - \frac{1}{\iu \hbar} \, H_N \,u(s)} = - \frac{a(s)}{\iu \hbar} \, \scal{\Phi^{(2)}\sub{gen}}{H\sub{int} \, \left[ \varphi_1(s) \wedge \cdots \wedge \varphi_N(s) \right]} \\
 & \quad = - \frac{a(s)}{\iu \hbar} \, \scal{\varphi_1(s) \wedge \cdots \wedge \theta_{m_1} \wedge \cdots \wedge \theta_{m_2} \wedge \cdots \wedge \varphi_N(s)}{H\sub{int} \, \left[ \varphi_1(s) \wedge \cdots \wedge \varphi_N(s) \right]} \,.
 \end{aligned}
 \end{equation}
 The above scalar product can be computed as in the proof of Theorem~\ref{thm:HF}, compare~\eqref{eqn:interac}, using again the orthogonality conditions between the $\theta$'s and the $\varphi$'s: one is lead to
 \begin{equation} \label{eqn:Phi(2)Vhat}
 \begin{aligned}
 & \scal{\varphi_1(s) \wedge \cdots \wedge \theta_{m_1} \wedge \cdots \wedge \theta_{m_2} \wedge \cdots \wedge \varphi_N(s)}{H\sub{int} \, \left[ \varphi_1(s) \wedge \cdots \wedge \varphi_N(s) \right]} \\
 & \quad = \scal{\theta_{m_1} \otimes \theta_{m_2}}{V_{12} \, \left[ \varphi_{m_1}(s) \otimes \varphi_{m_2}(s) -  \varphi_{m_2}(s) \otimes \varphi_{m_1}(s) \right]} \\
 & \quad = \sqrt{2} \, \scal{\theta_{m_1} \otimes \theta_{m_2}}{V_{12} \, \left[ \varphi_{m_1}(s) \wedge \varphi_{m_2}(s) \right]}\,.
 \end{aligned}
 \end{equation}
 
 Indeed, let us start from~\eqref{eqn:VijSlater}, dropping again for notational convenience the dependence on $s$. In analogy with~\eqref{Vij product}, we can write 
\begin{align*}
& \scal{\varphi_1 \wedge \cdots \wedge \theta_{m_1} \wedge \cdots \theta_{m_2} \wedge \cdots \wedge \varphi_N}{V_{ij}\,  \varphi_1 \wedge \cdots \wedge \varphi_N} \\
& \quad = \frac{1}{N!}\, \sum_{\sigma, \sigma' \in \mathfrak{S}_N} (-1)^{\sigma'} \, (-1)^\sigma \, \sum_{\alpha, \beta \in \N \times \Z/M\Z} v_{\alpha \beta} \, \\
& \qquad \scal{\varphi'_{\sigma'(1)} \otimes \cdots \otimes \varphi'_{\sigma'(N)}}{\varphi_{\sigma(1)} \otimes \cdots \otimes \left( \varphi_\alpha \, \varphi_{\sigma(i)} \right) \otimes \cdots \otimes \left( \varphi_\beta \, \varphi_{\sigma(j)} \right) \otimes \cdots \otimes \varphi_{\sigma(N)}} \\
& \quad = \frac{1}{N!}\, \sum_{\sigma, \sigma' \in \mathfrak{S}_N} (-1)^{\sigma'} \, (-1)^\sigma \, \sum_{\alpha, \beta \in \N \times \Z/M\Z} v_{\alpha \beta} \, \\
& \qquad \scal{\varphi'_{\sigma'(1)}}{\varphi_{\sigma(1)}} \, \cdots \, \scal{\varphi'_{\sigma'(i)}}{\varphi_\alpha \, \varphi_{\sigma(i)}} \, \cdots \, \scal{\varphi'_{\sigma'(j)}}{\varphi_\beta \, \varphi_{\sigma(j)}} \, \cdots \, \scal{\varphi'_{\sigma'(N)}}{\varphi_{\sigma(N)}}\,,
\end{align*} 
where for $\ell \in \set{1, \ldots, N}$
\[ \varphi'_{\ell} := \begin{cases} \varphi_{\ell} & \text{if } \ell \notin \set{m_1, m_2}\,, \\ \theta_\ell & \text{if } \ell \in \set{m_1,m_2}\,. \end{cases} \]
The only permutations which contribute to the sum are those for which $\set{\sigma'(i), \sigma'(j)} = \set{m_1,m_2}$ as sets. Let us fix $\alpha, \beta \in \N \times \Z / M \Z$ and look at one summand of the corresponding sum, for which we have that the only non-zero contributions are
\begin{align*}
& \sum_{\substack{\sigma' \in \mathfrak{S}_N \\ \sigma'(i) = m_1, \; \sigma'(j) = m_2}} (-1)^{\sigma'} \sum_{\sigma \in \mathfrak{S}_N} (-1)^\sigma \, \delta_{\sigma'(1), \, \sigma(1)} \, \cdots \, \scal{\theta_{m_1}}{\varphi_\alpha \, \varphi_{\sigma(i)}} \\
& \hspace{.45\textwidth} \cdots \, \scal{\theta_{m_2}}{\varphi_\beta \, \varphi_{\sigma(j)}} \, \cdots \, \delta_{\sigma'(N),\,\sigma(N)} \\[10pt]
& + \sum_{\substack{\sigma' \in \mathfrak{S}_N \\ \sigma'(i) = m_2, \; \sigma'(j) = m_1}} (-1)^{\sigma'} \sum_{\sigma \in \mathfrak{S}_N} (-1)^\sigma \, \delta_{\sigma'(1), \, \sigma(1)} \, \cdots \, \scal{\theta_{m_2}}{\varphi_\alpha \, \varphi_{\sigma(i)}}  \\
& \hspace{.45\textwidth} \cdots \, \scal{\theta_{m_1}}{\varphi_\beta \, \varphi_{\sigma(j)}} \, \cdots \, \delta_{\sigma'(N),\,\sigma(N)}\,.
\end{align*}

As in the proof of Theorem~\ref{thm:HF}, the only permutations $\sigma \in \mathfrak{S}_N$ which give a non-zero contribution to the sums above are $\sigma = \sigma'$ or $\sigma = \sigma' \circ (ij)$. Therefore, the above sums equal
\begin{align*}
& \sum_{\substack{\sigma' \in \mathfrak{S}_N \\ \sigma'(i) = m_1, \; \sigma'(j) = m_2}} \big[ \scal{\theta_{m_1}}{\varphi_\alpha \, \varphi_{m_1}} \, \cdot \, \scal{\theta_{m_2}}{\varphi_\beta \, \varphi_{m_2}} - \scal{\theta_{m_1}}{\varphi_\alpha \, \varphi_{m_2}} \, \cdot \, \scal{\theta_{m_2}}{\varphi_\beta \, \varphi_{m_1}} \big] \\
& + \sum_{\substack{\sigma' \in \mathfrak{S}_N \\ \sigma'(i) = m_2, \; \sigma'(j) = m_1}} \big[ \scal{\theta_{m_2}}{\varphi_\alpha \, \varphi_{m_2}} \, \cdot \, \scal{\theta_{m_1}}{\varphi_\beta \, \varphi_{m_1}} - \scal{\theta_{m_2}}{\varphi_\alpha \, \varphi_{m_1}} \, \cdot \, \scal{\theta_{m_1}}{\varphi_\beta \, \varphi_{m_2}} \big]\\
& = (N-2)! \cdot \scal{\theta_{m_1} \otimes \theta_{m_2}}{(\varphi_\alpha \otimes \varphi_\beta) \, \left[ \varphi_{m_1} \otimes \varphi_{m_2} - \varphi_{m_2} \otimes \varphi_{m_1} \right]} + \big( \beta \leftrightarrow \alpha \big)\,
\end{align*} 
as the summands on the left-hand side are independent of the permutation $\sigma' \in \mathfrak{S}_N$, and there are $(N-2)!$ such permutations which satisfy $\sigma'(i) = m_1$ and $\sigma'(j) = m_2$ and equally as many which satisfy $\sigma'(i) = m_2$ and $\sigma'(j) = m_1$. Using the symmetry~\eqref{eqn:valphabeta} of the potential, the term with $\beta$ and $\alpha$ swapped results in an overall factor of $2$, and we conclude that
\begin{align*}
& \scal{\varphi_1 \wedge \cdots \wedge \theta_{m_1} \wedge \cdots \theta_{m_2} \wedge \cdots \wedge \varphi_N}{V_{ij}\,  \varphi_1 \wedge \cdots \wedge \varphi_N} \\
 & \quad = \frac{2 \, (N-2)!}{N!} \, \scal{\theta_{m_1} \otimes \theta_{m_2}}{(\varphi_\alpha \otimes \varphi_\beta) \, \left[ \varphi_{m_1} \otimes \varphi_{m_2} - \varphi_{m_2} \otimes \varphi_{m_1} \right]} \\
 & \quad = \frac{2}{N\, (N-1)} \, \scal{\theta_{m_1} \otimes \theta_{m_2}}{(\varphi_\alpha \otimes \varphi_\beta) \, \left[ \varphi_{m_1} \otimes \varphi_{m_2} - \varphi_{m_2} \otimes \varphi_{m_1} \right]}
\end{align*} 
independently of $i < j \in \set{1, \ldots, N}$. Summing over the $\binom{N}{2}$ choices of such indices yields the desired conclusion, namely~\eqref{eqn:Phi(2)Vhat}.

An orthonormal basis for $\cH_N^{(2)}$ can be exhibited as follows: complete the orthonormal set ${\varphi_1(s), \ldots, \varphi_N(s)}$ to an orthonormal basis $\set{\varphi_1(s), \ldots, \varphi_N(s), \theta_1, \theta_2, \ldots}$ of $L^2(\Lambda)$. An orthonormal basis for $\cH_N^{(2)}$ is then provided by the vectors 
\[ \set{\Phi^{(2)}_{m_1,m_2;a_1,a_2} \colon 1 \le m_1 < m_2 \le N, \; 1 \le a_1 < a_2 < + \infty} \]
defined as
\begin{multline*}
\Phi^{(2)}_{m_1,m_2;a_1,a_2} \\
:= \varphi_1(s) \wedge \cdots \wedge \varphi_{m_1-1}(s)  \wedge \theta_{a_1} \wedge \varphi_{m_1+1}(s) \wedge \cdots \wedge \varphi_{m_2-1}(s) \wedge \theta_{a_2} \wedge \varphi_{m_2+1}(s) \wedge \cdots \wedge \varphi_N(s)\,. 
\end{multline*}

Let now $\Phi^{(2)}$ be any vector in $\cH_N^{(2)}$, which we decompose in the above basis as
\[ \Phi^{(2)} = \sum_{1 \le m_1 < m_2 \le N} \sum_{1 \le a_1 < a_2 < + \infty} \scal{\Phi^{(2)}_{m_1,m_2;a_1,a_2}}{\Phi^{(2)}} \, \Phi^{(2)}_{m_1,m_2;a_1,a_2}\, . \]
Then by~\eqref{eqn:scalPhi(2)} and~\eqref{eqn:Phi(2)Vhat} we can compute
\begin{align*}
\Biggl\langle \Phi^{(2)} & \,,\, \dot{u}(s) - \frac{1}{\iu \hbar} \, H_N \,u(s) \Biggr\rangle \\
&  = \sum_{1 \le m_1 < m_2 \le N} \sum_{1 \le a_1 < a_2 < + \infty} \overline{\scal{\Phi^{(2)}_{m_1,m_2;a_1,a_2}}{\Phi^{(2)}}} \\
& \hspace{.3\textwidth} \cdot \scal{\Phi^{(2)}_{m_1,m_2;a_1,a_2}}{\dot{u}(s) - \frac{1}{\iu \hbar} \, H_N \,u(s)} \\
& = - \frac{\sqrt{2} \, a(s)}{\iu \hbar} \, \sum_{1 \le m_1 < m_2 \le N} \sum_{1 \le a_1 < a_2 < + \infty} \overline{\scal{\Phi^{(2)}_{m_1,m_2;a_1,a_2}}{\Phi^{(2)}}} \\
& \hspace{.3\textwidth} \cdot 
\scal{\theta_{a_1} \otimes \theta_{a_2}}{V_{12} \, \left[ \varphi_{m_1}(s) \wedge \varphi_{m_2}(s) \right]}
\end{align*}
which we can bound as
\begin{align*}
& \left| \scal{\Phi^{(2)}}{\dot{u}(s) - \frac{1}{\iu \hbar} \, H_N \,u(s)} \right| \\
& \quad \le \frac{\sqrt{2}}{\hbar} \, \left( \sum_{1 \le m_1 < m_2 \le N} \sum_{1 \le a_1 < a_2 < + \infty} \left| \scal{\Phi^{(2)}_{m_1,m_2;a_1,a_2}}{\Phi^{(2)}} \right|^2 \right)^{1/2} \\
& \qquad \cdot 
\left( \sum_{1 \le m_1 < m_2 \le N} \sum_{1 \le a_1 < a_2 < + \infty} \left| \scal{\theta_{a_1} \otimes \theta_{a_2}}{V_{12} \, \left[ \varphi_{m_1}(s) \wedge \varphi_{m_2}(s) \right]} \right|^2 \right)^{1/2}\\
& \quad \le \frac{\sqrt{2}}{\hbar} \, \norm{\Phi^{(2)}} \, 
\left( \sum_{1 \le m_1 < m_2 \le N} \norm{\Pi \, V_{12} \, \left[ \varphi_{m_1}(s) \wedge \varphi_{m_2}(s) \right]}^2 \right)^{1/2}\,,
\end{align*}
where $\Pi$ is the projection onto the subspace of $L^2(\Lambda) \otimes L^2(\Lambda)$ spanned by the vectors $\set{\theta_{a_1} \otimes \theta_{a_2}}_{a_1, a_2 \in \N}$, obtained as tensor products of the functions orthogonal to all the Hartree--Fock orbitals $\varphi_j(s)$'s. On the other hand, as uniformly in $1 \le m_1 < m_2 \le N$ and in $s \in [0,t]$ we have
\[ \norm{\Pi \, V_{12} \, \left[ \varphi_{m_1}(s) \wedge \varphi_{m_2}(s) \right]} \le \norm{\Pi} \, \norm{V_{12}} \, \norm{\varphi_{m_1}(s) \wedge \varphi_{m_2}(s)} = \norm{V}_{L^\infty(\Lambda \times \Lambda)}\,, \]
we finally conclude that
\begin{equation} \label{eqn:EstimatePhi(2)}
\left| \scal{\Phi^{(2)}}{\dot{u}(s) - \frac{1}{\iu \hbar} \, H_N \,u(s)} \right| \le \frac{1}{\hbar} \, \norm{\Phi^{(2)}} \, \norm{V}_{L^\infty(\Lambda \times \Lambda)} \, \sqrt{N(N-1)}\,,
\end{equation}
for all $\Phi^{(2)} \in \cH_N^{(2)}$.

\item For $\Phi^{(i)}\sub{gen}$ a generating Slater determinant in the subspace $\cH_N^{(i)}$ with $i \ge 3$, we can see that
 \begin{gather*}
 \scal{\Phi^{(i)}\sub{gen}}{u(s)} = a(s) \, \scal{\Phi^{(i)}\sub{gen}}{\varphi_1(s) \wedge \cdots \wedge \varphi_N(s)} = 0\,, \\
 \scal{\Phi^{(i)}\sub{gen}}{\varphi_1(s) \wedge \cdots \wedge \eta_\ell(s) \wedge \cdots \wedge \varphi_N(s)} = 0\,, \\
 \scal{\Phi^{(i)}}{H\sub{int} \, \left[ \varphi_1(s) \wedge \cdots \wedge \varphi_N(s) \right]} = 0\,.
 \end{gather*}
The only non-trivial identity is the third, which can be shown via a similar argument to the one just presented, starting from~\eqref{eqn:VijSlater}: indeed, the scalar product $\scal{\Phi^{(i)}\sub{gen}}{V_{ij} \, \left[ \varphi_1(s) \wedge \cdots \wedge \varphi_N(s) \right]}$ can be decomposed as a sum of products, each of which contains at least a factor of the form $\scal{\theta_m}{\varphi_\ell} = 0$. We conclude that 
\[ \scal{\Phi^{(i)}}{\dot{u}(s) - \frac{1}{\iu \hbar} \, H_N \,u(s)} = 0 \quad \text{for all } \Phi^{(i)} \in \cH_N^{(i)}, \; i \in \set{3, \ldots, N}\,. \]
\end{itemize}

Summing up the above considerations we have deduced that
\[ \scal{\Phi}{\dot{u}(s) - \frac{1}{\iu \hbar} \, H_N \,u(s)} = \sum_{i=0}^{N} \scal{\Phi^{(i)}}{\dot{u}(s) - \frac{1}{\iu \hbar} \, H_N \,u(s)} = \scal{\Phi^{(2)}}{\dot{u}(s) - \frac{1}{\iu \hbar} \, H_N \,u(s)} \]
which, coming back to~\eqref{eqn:dualnorm}, together with~\eqref{eqn:EstimatePhi(2)} yields
\begin{align*}
\norm{\dot{u}(s) - \frac{1}{\iu \hbar} \, H_N \,u(s)} & = \sup_{\substack{\Phi \in \cH_N \\ \norm{\Phi} = 1}} \left| \scal{\Phi}{\dot{u}(s) - \frac{1}{\iu \hbar} \, H_N \,u(s)} \right| \\
& = \sup_{\substack{\Phi^{(2)} \in \cH_N^{(2)} \\ \norm{\Phi^{(2)}} = 1}} \left| \scal{\Phi^{(2)}}{\dot{u}(s) - \frac{1}{\iu \hbar} \, H_N \,u(s)} \right| \\
& \le \frac{1}{\hbar} \, \sqrt{N(N-1)} \, \norm{V}_{L^\infty(\Lambda \times \Lambda)} 
\end{align*}
uniformly in $s \in [0,t]$. Integrating over this interval, we conclude together with~\eqref{eqn:APosteriori} that for all $t \in [0,\overline{t}]$
\[ \norm{\Psi(t) - u(t)} \le \frac{1}{\hbar} \, \sqrt{N(N-1)} \, \norm{V}_{L^\infty(\Lambda \times \Lambda)} \, t \]
as claimed.
\end{proof}

\begin{remark}
In the above proof, \eqref{eqn:HF0thorder} and~\eqref{eqn:HF1storder} show once more that the choice of the nonlinear part of the Hartree--Fock equations determines a partial cancellation of the effects of the interacting part $H\sub{int}$ of the Hamiltonian, when the vector in~\eqref{eqn:udot-Hu} is projected on Slater determinants where at most one orbital is orthogonal to those evolving according to the effective dynamics. The error terms come instead from pair interactions with pairs of orbitals both orthogonal to those satisfying the Hartree--Fock equations, compare~\eqref{eqn:Phi(2)Vhat}. This argument is quite general and abstract (compare \cite{Lubich08}), and does not use the specific form of the non-interacting part of the Hamiltonian $H_N$ in an essential way: indeed, as briefly mentioned in the Introduction, the same argument could be adopted in the non-magnetic case, namely replacing the Landau Hamiltonian with the free one at the one-body level.
\end{remark}

We end with some comments on the effectiveness of the estimate~\eqref{eqn:Estimate}. The approximation of the Schr\"odinger dynamics by a Slater determinant by definition ignores all effects of correlations contained in the many-body wave function due to the interaction. As expected, initializing the many-body dynamics at an uncorrelated state, the effective dynamics is supposed to give a sufficiently good approximation of this wave function \emph{for small times} --- hence the linear dependence on $t$ on the right-hand side of~\eqref{eqn:Estimate} --- and \emph{for weak interactions} --- as indicated by the dependence on $\norm{V}_{L^\infty(\Lambda \times \Lambda)}$ of the bound. The dependence of the bound~\eqref{eqn:Estimate} on the number of particles is instead essentially \emph{linear}, proportional to $N$. The bound in~\eqref{eqn:Estimate} should be compared to the trivial bound $\norm{\Psi(t) - u(t)} \le 2$ provided by the triangle inequality: with this proof, the nonlinear dynamics appears to be effective then for very small times, of order $t \ll 1/N$.

To put this consideration in perspective, it is convenient to take into account both the energy and time scales. From Remark~\ref{rmk:E0N} and~\eqref{eqn:normhatV}, we see that, at least in non-interacting ground states, both the kinetic, non-interacting energy and the two-body interactions scale quadratically in the number of particles. This suggests that, in the large-$N$ limit, each particle experiences an ``averaged effect'' of the interaction with all the other particles, that is, that the interaction potential is represented by the effects of a \emph{mean field} generated by the other $N-1$ particles. In non-magnetic fermionic systems, it is common (see e.g.\ the discussion in~\cite{BenedikterPortaSchlein16, BenedikterDesio23}) to also couple the large-$N$ limit to an appropriate \emph{semiclassical} scaling, using a rescaled effective Planck constant. In practice, these coupled mean-field and semiclassical limits require rescaling
\[ \hbar \leadsto \hbar\sub{eff} := \hbar \, N^{-1/2}, \quad V(\vec{x};\vec{y}) \leadsto V\sub{eff}(\vec{x};\vec{y}) := \frac{1}{N} \, V(\vec{x};\vec{y})\,. \]
Upon this rescaling, the bound~\eqref{eqn:Estimate} becomes
\begin{align*}
\norm{\Psi(t) - u(t)}_{\cH_N} & \le \frac{1}{\hbar\sub{eff}} \, \sqrt{N(N-1)} \, \norm{V\sub{eff}}_{L^\infty(\Lambda \times \Lambda)} \, t \\
& = \frac{\sqrt{N}}{\hbar} \, \sqrt{N} \, \sqrt{N-1} \, \frac{1}{N} \, \norm{V}_{L^\infty(\Lambda \times \Lambda)} \, t \\
& = \frac{1}{\hbar} \, \sqrt{N-1} \, \norm{V}_{L^\infty(\Lambda \times \Lambda)} \, t\,,
\end{align*}
which shows that the nonlinear Hartree--Fock dynamics is an effective description of the Schr\"odinger evolution, in the coupled mean-field and semiclassical scalings, for times smaller than the semiclassical scale: $t \ll 1/\sqrt{N} = N^{-1/2}$.

Again in absence of magnetic fields, a much more refined control on the interaction energy allows to improve the $N$-dependence of the bound, at the expense of a super-exponential (rather than linear) dependence on time. This approach was pursued in~\cite{BenedikterPortaSchlein14, PetratPickl16, BenedikterDesio23, ChenLeeLiLiew23} for the coupled mean-field and semiclassical scaling presented above, and in~\cite{FrestaPortaSchlein23} in the so-called \emph{Kac regime}. The comparison is formulated by introducing the \emph{$1$-body reduced density matrices} corresponding to the states $\Psi(t)$ and $u(t)$, defined respectively as the operators on $L^2(\Lambda)$ whose integral kernels are given by~\cite{LiebSeiringer09}
\begin{gather*}
\omega_{\Psi}(t, \vec{x};\vec{y}) := N \, \int_{\Lambda^{N-1}} \di \vec{x}_2 \ldots \di \vec{x}_N \, \Psi(t,\vec{x}, \vec{x}_2, \ldots, \vec{x}_N) \, \overline{\Psi(t,\vec{y}, \vec{x}_2, \ldots, \vec{x}_N)}\,, \\
\omega\sub{Slater}(t, \vec{x};\vec{y}) := \sum_{\ell=1}^{N} \varphi_\ell(t,\vec{x}) \, \overline{\varphi_\ell(t,\vec{y})}\,.
\end{gather*}
The operators $\omega_{\Psi}(t)$ and $\omega\sub{Slater}(t)$ are then non-negative trace-class operators, with trace equal to $N$ --- in fact, $\omega\sub{Slater}(t)$ is the rank-$N$ projection onto the subspace of $L^2(\Lambda)$ spanned by the orthonormal orbitals $\varphi_1(t), \ldots, \varphi_N(t)$. The difference $\omega_{\Psi}(t) - \omega\sub{Slater}(t)$ can be then estimated in trace norm: while the triangle inequality would give a bound of order $N$, a much smaller error of order $\sqrt{N}$ can be achieved. The generalization of these methods to systems of fermions in a magnetic field is certainly an interesting future line of research.

\subsection*{Acknowledgements}

We are thankful to Niels Benedikter, Chiara Boccato, Emanuela Giacomelli and Gianluca Panati for discussions related to the topics presented in this note.

M.~F.~gratefully acknowledges financial support from DFG--German Research Foundation within CRC TRR 352 ``Mathematics of Many-Body Quantum Systems and Their Collective Phenomena'' (Project-ID 470903074). D.~M.~gratefully acknowledges financial support from Sapienza Universit\`{a} di Roma within Progetto di Ricerca di Ateneo 2020, 2021, 2022 and 2023 and from MUR--Italian Ministry of University and Research and Next Generation EU within PRIN 2022AKRC5P ``Interacting Quantum Systems: Topological Phenomena and Effective Theories''. 

This work has been carried out under the auspices of the GNFM-INdAM (Gruppo Nazionale per la Fisica Matematica --- Istituto Nazionale di Alta Matematica) and within the framework of the activities for PNRR MUR under Project No. PE0000023-NQSTI.

\label{lastpage}
\end{document}